\documentclass{article}
\pdfoutput=1
\usepackage{times,fancyhdr}
\usepackage{cite}
\usepackage{graphicx,amsmath,amssymb}

\DeclareMathAlphabet{\mathpzc}{OT1}{pzc}{m}{it}

\newcommand{\ket}[1]{|#1\rangle}

\title{A semi-infinite matrix analysis of the BFKL equation 
} 
\author{N. Bethencourt de Le{\'o}n$^1$, G. Chachamis$^2$, A. Romagnoni$^{3,4}$,  A. Sabio Vera$^{1,5}$ \\ 
\\
$^1$ Instituto de F{\' \i}sica Te{\' o}rica UAM/CSIC, Nicol{\'a}s Cabrera 15, \\ 
\& Universidad Aut{\' o}noma de Madrid, E-28049 Madrid, Spain. \\
$^2$ LIP, Av. Prof. Gama Pinto, 2, P-1649-003 Lisboa, Portugal.\\
 $^3$ Centre de recherche sur l’inflammation UMR 1149,\\ 
Inserm - Universit{\'e} Paris Diderot - 75018 Paris, France.\\
$^4$Data Team, D{\' e}partement d'informatique de l’ENS, {\'E}cole normale sup{\'e}rieure,\\ CNRS, PSL Research University, 75005 Paris, France.\\
$^5$ CERN, Theoretical Physics Department, Geneva, Switzerland. 
} 

\begin{document} 

%\pagestyle{fancy}
%\fancyhead{}
%\fancyhead[EC]{Martin Hentschinski \& Agust{\' \i}n Sabio Vera}
%\fancyhead[EL,OR]{\thepage}
%\fancyhead[OC]{NLO jet vertex from Lipatov's high energy effective action in QCD}
%\fancyfoot{} 
%\renewcommand\headrulewidth{0.5pt}
%\addtolength{\headheight}{2pt} 

\maketitle 

The forward BFKL equation is discretised in virtuality space and it is shown that the diffusion into infrared and ultraviolet momenta can be understood in terms of a semi-infinite matrix. The square truncation of this 
matrix can be exponentiated leading to asymptotic eigenstates sharing many features with the BFKL gluon Green's function in the limit of large matrix size. This truncation is closely related to a representation of the XXX Heisenberg  spin~$= - \frac{1}{2}$ chain with SL(2) invariance where the Hamiltonian  acts on a symmetric double copy of the harmonic oscillator. 
A simple modification of the BFKL matrix suppressing the infrared modes generates evolution more compatible with the Froissart bound. 

\section{Introduction}

In recent years there has been a growing activity concerning the identification of integrable structures in four-dimensional gauge theories. This is mainly due to the interest that this subject has for the 
anti de Sitter / conformal field theory (AdS/CFT) conjecture~\cite{Maldacena}. 
After the seminal works in~\cite{Minahan:2002ve}, big progress has been made in the mapping of anomalous dimensions of gauge invariant Wilson operators in super Yang-Mills (SYM) theory to the spectrum of string theory in different backgrounds. A crucial step was to realize that the planar one-loop dilatation operator of ${\cal N} = 4$ SYM 
maps into the Hamiltonian of an integrable spin chain. The problem of finding anomalous dimensions translates then into the diagonalization of the corresponding Hamiltonian, and all the techniques developed for integrable systems become very 
useful in accomplishing this task. After those first results, the better understanding of the mapping has allowed more general results for larger orders in perturbation theory and for different sectors of the gauge and string theories (for an introduction to 
the field and a wider bibliography see~\cite{Beisert:2010jr}).

Nonetheless, two-dimensional integrable structures in four-dimensional gauge field theory appeared well before the AdS/CFT conjecture, in the region of high energy scattering in Quantum Chromodynamics (QCD). Non-Abelian gauge theories manifest interesting mathematical properties when they are investigated in terms of high energy scattering amplitudes 
in the Regge limit. This is the case of the SL(2,$\mathbb{C}$) invariance~\cite{Lipatov:1985uk}  present in the impact parameter representation of QCD (and ${\cal N} = 4$ SYM) elastic 
scattering amplitudes evaluated in multi-Regge 
kinematics~\cite{Lipatov:1976zz,Fadin:1975cb,Kuraev:1976ge,Kuraev:1977fs,Balitsky:1978ic}. In this context the 
Balitsky-Fadin-Kuraev-Lipatov (BFKL) pomeron (with vacuum quantum numbers exchanged in the $t$-channel) can be interpreted as a bound state of two reggeized gluons where the Hamiltonian has an interesting operator representation~\cite{Lipatov:1996ts} with holomorphic separability in coordinate space~\cite{Lipatov:1990zb}. The iteration of the BFKL 
Hamiltonian in the $s$-channel, describing multiple reggeon exchanges in the generalized leading 
logarithmic approximation, defines the Bartels-Kwiecinski-Praszalowicz (BKP) 
equation~\cite{Bartels:1980pe,Kwiecinski:1980wb} and was found to have a hidden 
integrability~\cite{Lipatov:1985uk,Lipatov:1990zb,Lipatov:1993qn,Lev1}, being equivalent to a periodic spin chain of a XXX Heisenberg 
ferromagnet~\cite{Lev2,Lipatov:1994xy,Faddeev:1994zg}. 
This was the first example of the existence of integrable systems in QCD. 
A similar integrable spin chain, an open one this time, was found in kinematical regions of $n$-point 
maximally helicity violating (MHV) and planar ($N_c \to \infty$) amplitudes in  ${\cal N}=4$ SYM where Mandelstam cut contributions are maximally enhanced~\cite{Lipatov:2009nt}. The importance of Mandelstam cuts in the 
complex angular momentum plane for ${\cal N}=4$ SYM MHV planar amplitudes was first realized in~\cite{Bartels:2008ce,Bartels:2008sc} 
where corrections to the Bern-Dixon-Smirnov (BDS) iterative ansatz~\cite{Bern:2005iz} for this class of amplitudes were found for the six-point amplitude at two loops.

There is an interesting connection between the integrable structures appearing in the calculation of the anomalous dimension of gauge invariant twist (scaling dimension minus Lorentz spin) two operators of spin $M$ in ${\cal N}=4$ SYM, and in the multi-Regge kinematics. As it was shown in\cite{Dressing}, the link to the BFKL equation appears upon analytically continuing the anomalous dimension function to complex values of $M$. In particular, the pomeron corresponds to the first singularity at $M = \omega -1$, for small $\omega$. The discrepancy between this result and the prediction obtained from  the asymptotic Bethe ansatz was subsequently explained in~\cite{Janik} by the calculation of the corresponding wrapping corrections for the twist two operators. 

The aim of the present work is to introduce a formal representation of the BFKL equation in matrix form which allows to investigate it in momentum space in a novel way. This framework is flexible enough to allow for modifications which tame the growth with energy of the BFKL evolution and which can be interpreted as an absorptive barrier for infrared modes. 
It is then shown how this approach, truncated in the ultraviolet, is closely related to the two-sites one-loop Hamiltonian of the sl(2) sector of the ${\cal N}=4$ SYM theory, in the double oscillator picture for operators of a given spin~\cite{Beisert:2003jj}. 

After this brief Introduction to the subject, in Section 2, a general discussion on the BFKL equation is provided, explaining the connection between the non-forward and forward limits. A novel discretization in virtuality space 
is described in detail, highlighting the role and physical interpretation of the shift and diagonal operators 
appearing in this representation. In Section 3 a square truncation of the BFKL discretization is introduced  and  the asymptotic behaviour of the corresponding eigensystem investigated. Section 4 is devoted to the study of a modification of the matrix Hamiltonian which reduces the influence of propagation into the infrared. In Section 5, known facts about Beisert's representation of the non-compact SL(2) spin chain are introduced to set the ground for a comparison with the BFKL equation. Finally, some Conclusions are drawn. 

\section{Matrix representation}

In this work the BFKL Hamiltonian is considered directly in two-dimensional transverse momentum space ($\vec{k}$), where other components have decoupled into an evolution variable (rapidity $Y$) which plays the role of 
time~\cite{Lipatov:1976zz,Fadin:1975cb,Kuraev:1976ge,Kuraev:1977fs,Balitsky:1978ic}. The four-point amplitude for 
off-shell reggeized gluons has the following momentum flow:
\begin{center}
\includegraphics[height = 3cm]{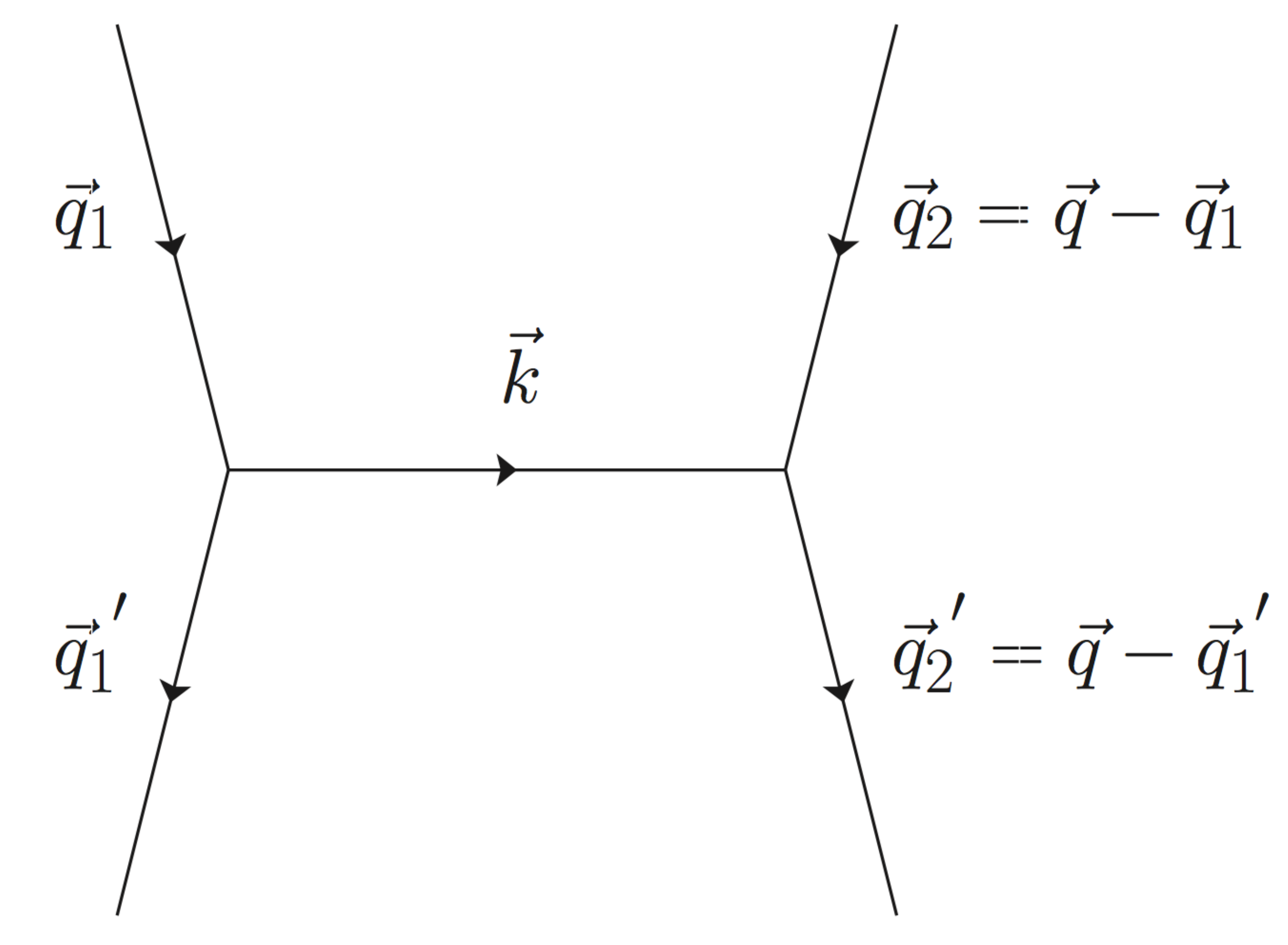}
\end{center}
With this notation, the BFKL kernel for the pomeron channel has two contributions. The first one corresponds to 
$\vec{k} = 0$, {\it i.e.}, there is no propagator in the $s$-channel and can be written as
\begin{eqnarray}
``{\rm Reggeized ~Propagators}" &\simeq& g^2 N_c \delta^{(2)} \left(\vec{q}_1 - \vec{q}_1^{~'}\right)
\delta^{(2)} \left(\vec{q}_2 - \vec{q}_2^{~'}\right) \nonumber\\
&& \hspace{1cm} \times \left(\int d^2 \vec{r} \frac{\vec{q}_1^{~2}}{\vec{r}^{~2} (\vec{q}_1 - \vec{r})^2}
+ \int d^2 \vec{r} \frac{\vec{q}_2^{~2}}{\vec{r}^{~2} (\vec{q}_2 - \vec{r})^2}\right).
\end{eqnarray}
The second piece has $\vec{k} \neq 0$ and corresponds to squaring the Lipatov's vertex, {\it i.e},
\begin{eqnarray}
``{\rm Emission}" &\simeq& \delta^{(2)} \left(\vec{q}_1 + \vec{q}_2 - \vec{q}_1^{~'} - \vec{q}_2^{~'}\right) 
\frac{g^2 N_c }{\vec{q}_1^{~2} \vec{q}_1^{~'2}} 
\left(
\frac{\vec{q}_1^{~2} \vec{q}_2^{~'2}+\vec{q}_2^{~2} \vec{q}_1^{~'2}}{\vec{k}^{ 2}} - \left(\vec{q}_1 + \vec{q}_2 \right)^2 
\right).
\end{eqnarray}
After a Fourier transform of these two expressions, and complexifying the transverse momenta, Lipatov found 
the SL(2,$\mathbb{C}$) invariance of this Hamiltonian~\cite{Lipatov:1985uk}. In the present work, however, the focus lies on the forward limit, with zero momentum transfer $\vec{q} = 0$. It is noteworthy that in this case the contribution from the ``Reggeized Propagators" reads
\begin{eqnarray}
``{\rm Reggeized ~Propagators}" &\simeq& 2 g^2 N_c \delta^{(2)} \left(\vec{q}_1 - \vec{q}_1^{~'}\right)
 \int d^2 \vec{r} \frac{\vec{q}_1^{~2}}{\vec{r}^{~2} (\vec{q}_1 - \vec{r})^2},
\label{reggeizedpropagators}
\end{eqnarray}
while the ``Emission" piece simplifies to
\begin{eqnarray}
``{\rm Emission}" &\simeq& 2 \frac{g^2 N_c }{\vec{k}^{2}} 
~=~ 2 \frac{g^2 N_c }{(\vec{q}_1 - \vec{q}_1^{~'})^{2}}.
\label{emission}
\end{eqnarray}
It is in this forward case that it truly represents a real emission since now the amplitude corresponds 
to the $2 \to 3$ inelastic process. A further simplification is very convenient: to integrate over the azimuthal angle 
formed by the two transverse momenta $\vec{q}_1$ and $\vec{q}_1^{~'}$. 

Once this is done the BFKL equation for forward scattering can be cast in the simple form
\begin{eqnarray}
{\partial \varphi (Q^2,Y) \over \alpha \partial Y } &=& 
\int_0^\infty {d q^2 \over |q^2-Q^2| } \left(\varphi (q^2,Y) - 
{2 \, {\rm min} (q^2,Q^2) \over q^2 + Q^2 } \varphi(Q^2,Y)  \right),
\label{eqn-q5}
\end{eqnarray}
where $\alpha = \alpha_s N_c / \pi$, the integration takes place over the gluon virtuality 
$q^2 (\equiv\vec{q}^{~2})$ and the correspondence with the previous notation is $\vec{q}_1^{~2} = Q^2$ and $\vec{q}_1^{~'2} = Q_0^2$. The term with $\varphi(Q^2,Y)$ 
corresponds to Eq.~(\ref{reggeizedpropagators}) and the one with $\varphi (q^2,Y)$ to Eq.~(\ref{emission}). 
Since the forward limit has been taken, $\varphi(Q^2,Y)$ is the cut reggeized gluon four-point function for a given rapidity $Y$ with the initial condition $\varphi(Q^2,Y=0) \sim \delta(Q^2-Q_0^2)$, and it corresponds to the sum of the squares of the $2 \to 2 + n$ emissions amplitude over any number $n$ of real gluon emissions. 

To find the gluon Green's function it is convenient to write Eq.~(\ref{eqn-q5}) in the form  
\begin{eqnarray}
{\partial \varphi (Q^2,Y) \over \alpha \partial Y } &=&
\int_0^1 {d x \over 1-x } \left( \varphi (x \, Q^2,Y) + \frac{1}{x} 
\varphi \left( \frac{Q^2}{x},Y\right)- 2 \, \varphi(Q^2,Y)  \right) ,
\label{eqn-x}
\end{eqnarray}
and then introduce a Mellin transform to obtain
\begin{eqnarray}
\varphi (Q^2,Y) &=& \int_{a - i \infty}^{a+i \infty} \frac{d \gamma}{ 2 \pi i} 
\left({Q^2 \over Q_0^2}\right)^{\gamma-1} e^{\alpha Y \chi(\gamma)}, 
\label{Mellin}\\
\chi(\gamma) &=&  2 \psi(1) - \psi(\gamma) - \psi (1-\gamma),
\label{chi}
\end{eqnarray}
with $\psi$ being the digamma function and $0<a<1$. It is well-known that, for asymptotically 
large values of the rapidity variable $Y$, this integral tends to 
\begin{eqnarray}
\varphi (Q^2,Y) &\simeq& {e^{\alpha Y 4 \log{2}-\frac{t}{2} 
-{t^2 \over 56 \zeta(3) \alpha Y}}
\over 2 \sqrt{14 \pi \zeta(3) \alpha Y}},
\label{asymptotic}
\end{eqnarray}  
with $t=\log{Q^2/Q_0^2}$.  This implies the following diffusion 
equation for the function $\phi = \varphi e^{t/2} / \pi$:
\begin{eqnarray}
{\partial \phi \over \alpha \partial Y} &=& 4 \log{2} \,  \phi + 14 \zeta(3) 
{\partial^2 \phi \over \partial t^2},
\end{eqnarray}
which shows that there exist two different flows 
for the virtualities of the $t$-channel gluons in the BFKL ladder: one towards the 
infrared (IR) and one towards the ultraviolet (UV). These IR/UV flows are symmetric since the 
eigenvalue function $\chi (\gamma)$ is invariant under $\gamma \leftrightarrow 1- \gamma$. 
   
The space of virtualities can be discretized in  Eq.~(\ref{eqn-q5}) using $q^2=n \,\Delta$, $Q^2 = N \,\Delta$, 
$dl^2 = \Delta$ and the notation $\phi_n \equiv \varphi(n \,\Delta,Y)$. It is then possible to write (with $N= 1, \dots , \infty$)
\begin{eqnarray}
\frac{\partial \phi_N}{\alpha \partial Y} &=& \sum_{n=1}^{N-1} \frac{1}{N -n }\left(\phi_n  - \frac{2 n }{N +n}
\phi_N \right) +\sum_{n=N+1}^\infty \frac{1}{n - N}
\left( \phi_n - \frac{2 N}{N + n}\phi_N\right) \nonumber\\
&\simeq& \sum_{n=1}^{N-1} \frac{\phi_n }{N -n }+\sum_{n=N+1}^\infty \frac{\phi_n }{n-N }- 2 h(N-1)  \phi_N,
\label{maineq11}
\end{eqnarray}
where $h (N) = \sum_{l=1}^N \frac{1}{l} = \psi(N+1)- \psi (1)$ is the harmonic number. 
This is a valid representation up to ${\cal O} \left({\phi_N \over N }\right)$ terms, which are negligible  
at large $N$.

\begin{figure}[htbp]
  \centering
  \vspace{-.5cm}
  \includegraphics[width=8cm,angle=0]{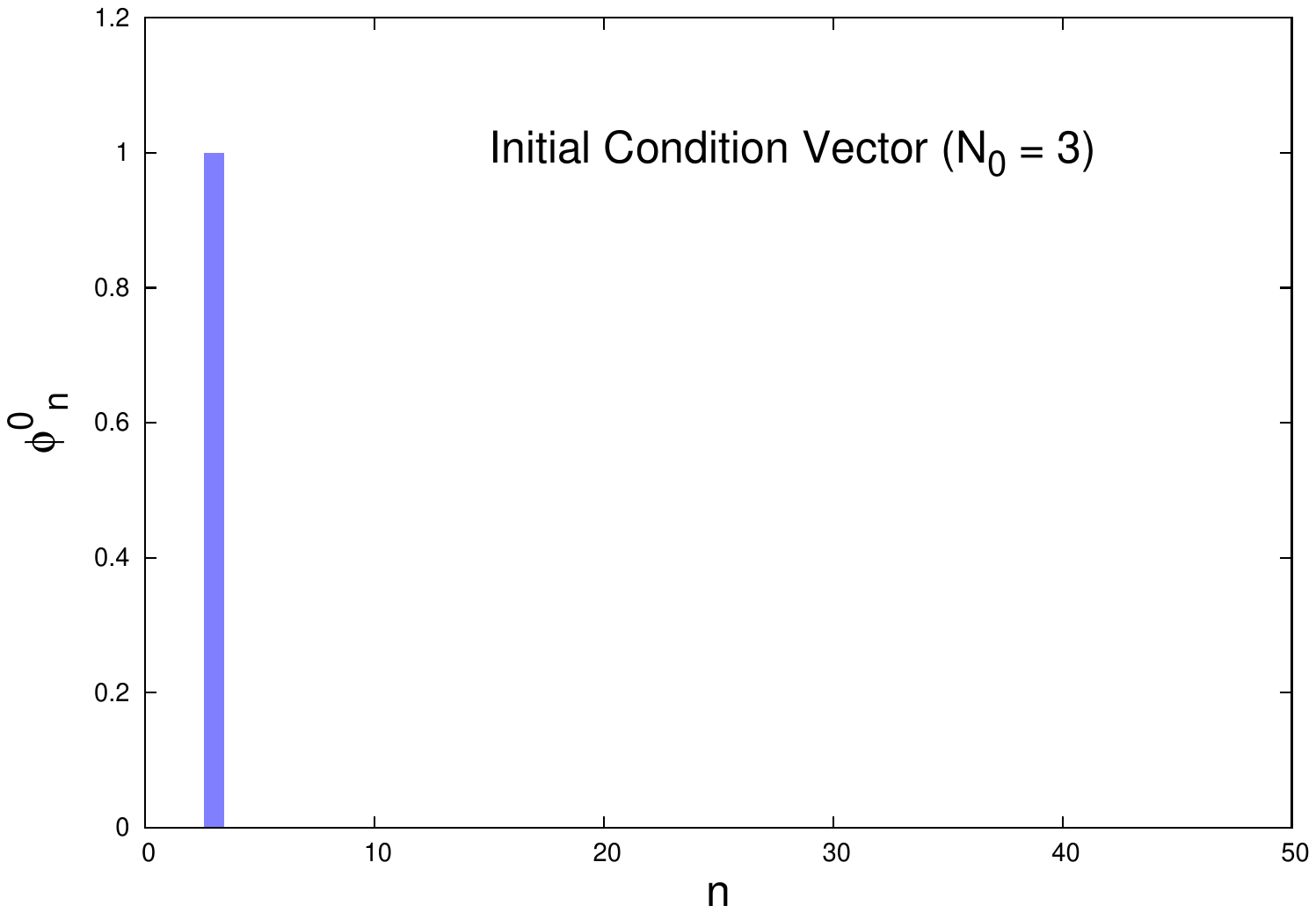}
  \vspace{-1.cm}
  \caption{Initial condition vector $\phi^0_i = \frac{\delta_i^{N_0}}{\Delta}$ taking $N_0 =3$ for $\Delta = 1$ and $N=50$.}
  \label{InitialCondition}
\end{figure}

\begin{figure}[htbp]
  \centering
  \vspace{-.5cm}
  \includegraphics[width=8cm,angle=0]{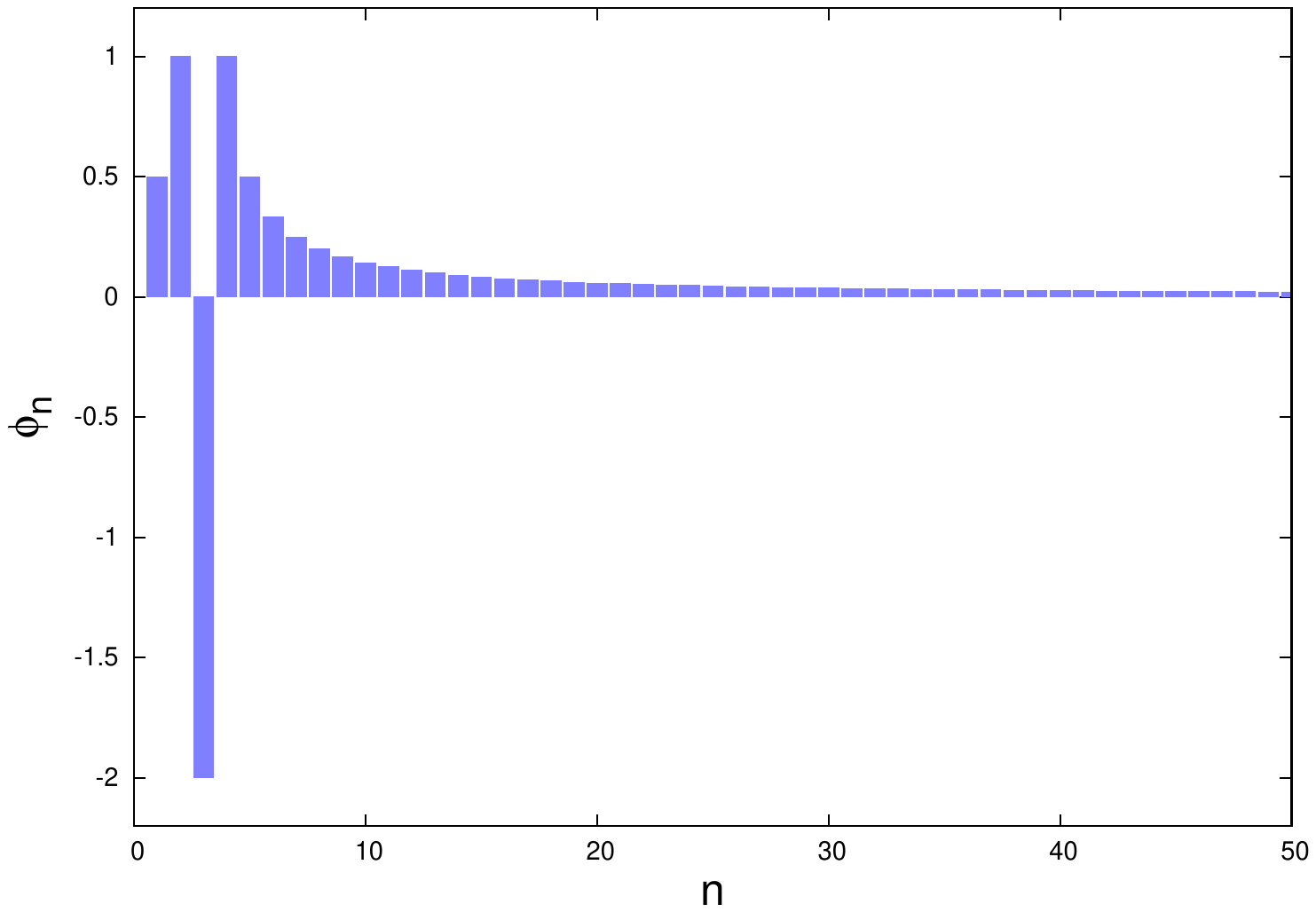}
  \vspace{-1.cm}
  \caption{Action of $\hat{\cal H}$ on $\phi^0_i = \frac{\delta_i^{N_0}}{\Delta}$ with $N=50$  and $N_0=3$.}
  \label{1actionofH}
\end{figure}

To find a matrix representation for the action of the kernel it is useful to introduce the $N$-dimensional vector 
$\vec{\phi} \equiv (\phi_1,\phi_2, \dots, \phi_N)^t$, the extended 
$\infty$-dimensional vector $\vec{\phi}_\infty \equiv (\phi_1,\phi_2, \dots, \phi_N, \dots)^t$ and write Eq.~(\ref{maineq11}) in the form
\begin{eqnarray}
\frac{\partial \vec{\phi}}{\alpha \partial Y} &=& 
\hat{\cal H} \cdot \vec{\phi}_\infty,
\label{eqn-short12}
\end{eqnarray}
with the kernel being the following semi-infinite matrix with $N$ rows and 
$\infty$ columns:
\begin{eqnarray}
\hat{\cal H} &=&
\left(\begin{array}{ccccccc}
-2 h(0) & 1 & \frac{1}{2} & \frac{1}{3}& \frac{1}{4} & \frac{1}{5}&\cdots\\
1& -2 h(1) & 1 & \frac{1}{2} & \frac{1}{3}& \frac{1}{4} &\cdots\\
\frac{1}{2}&1& - 2 h(2) & 1 & \frac{1}{2} & \frac{1}{3}& \cdots\\
\vdots&\vdots&\vdots& \vdots & \vdots & \vdots & \vdots\\
\frac{1}{N-1}&\frac{1}{N-2}&\cdots&1& -2 h(N-1) & 1 & \cdots\\
\end{array}\right). 
\label{matrixH}
\end{eqnarray}
In terms of components this is equivalent to 
\begin{eqnarray}
\left(\hat{\cal H}\right)_{i,j} &=& 
\sum_{n=1}^{N-1} \frac{\delta_i^{j+n}}{n} +
\sum_{n=1}^\infty \frac{\delta_{i+n}^j}{n} - 2 h(i-1) \delta_i^j,  
\end{eqnarray}
for $1 \leq i \leq N, 1 \leq j < \infty$. 
These matrix elements can be written in terms of the following shift operators:
\begin{eqnarray}
\left(\hat{\cal S}_{\rm IR}\right)_{i,j} = \delta_i^{j+1} \, \, \, , \, \, \, 
\left(\hat{\cal S}_{\rm UV}\right)_{i,j} = \delta_{i+1}^j.
\end{eqnarray}
Using the notation $\left(\hat{\cal G} \right)_{i,j} = - 2 h(i-1) \delta_i^j$, the Hamiltonian becomes
\begin{eqnarray}
\hat{\cal H} &=& \sum_{n=1}^{N-1} \frac{(\hat{\cal S}_{\rm IR})^n}{n} +
\sum_{n=1}^\infty \frac{(\hat{\cal S}_{\rm UV})^n}{n}+ \hat{\cal G}, 
\label{Ston}\\
&=& - \sum_{n=N}^\infty \frac{\hat{\cal S}_{\rm IR}^n}{n} - 
\log{\left(1-\hat{\cal S}_{\rm IR}\right)} -
\log{\left(1-\hat{\cal S}_{\rm UV}\right)} + \hat{\cal G}.
\label{logS}
\end{eqnarray}
The  diffusion picture is then related to the action of this matrix on an initial condition of the form $\sim \delta (Q^2-Q_0^2)$.  Since $Q_0^2 = N_0 \, \Delta$, the delta function corresponds to a single entry in the initial condition vector, {\it i.e.}
\begin{eqnarray}
\vec{\phi}_0 \equiv (\phi^0_1,\phi^0_2, \dots, \phi^0_N, \dots)^t \, \, \,  \, \, \, 
{\rm with} \, \, \, \, \, \, \phi^0_i = \frac{\delta_i^{N_0}}{\Delta}.
\end{eqnarray}
The action of the shift operators in the kernel translates the original
single ``cell'' $\phi^0_i$ towards lower or higher entries in 
the vector. This corresponds to
the emission of real gluons in the $s$-channel, which modify the virtuality 
of the $t$-channel reggeized gluons. The diagonal
piece in the kernel  corresponds to the generation of a rapidity gap in between gluon emissions. 
In order to illustrate this point, a matrix size $N=50$ has been chosen and  the matrix $\hat{\cal H}$ has been applied to the initial condition vector $\vec{\phi}_0$ with $N_0=3$ shown in Fig.~\ref{InitialCondition} ($\Delta=1$ has been taken).
The diffusion pattern  can be seen in the resulting components of 
$\hat{\cal H} \cdot \vec{\phi}_0$ in Fig.~\ref{1actionofH}.

\begin{figure}[htbp]
  \centering
  \vspace{-.5cm}
  \includegraphics[width=8cm,angle=0]{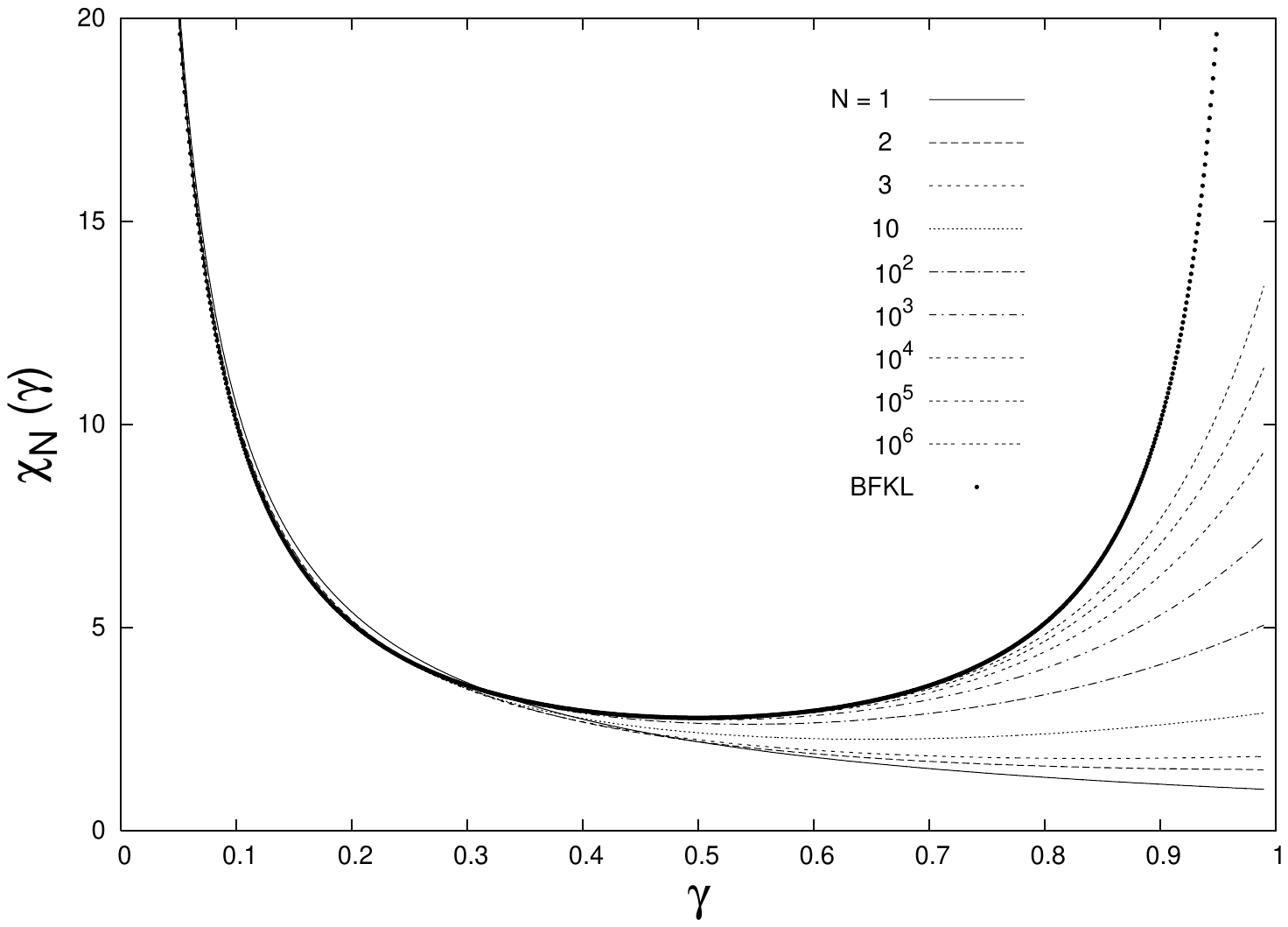}
  \vspace{-1.cm}
  \caption{$\chi_N (\gamma)$ coincides with the BFKL eigenvalue $\chi(\gamma)$ at $N \to \infty$. }
  \label{Kernels}
\end{figure}

The continuum limit corresponds to $N \to \infty$, $\Delta \to 0$, 
while keeping $N \Delta = Q^2$ fixed. In order to investigate this point in more detail one can rewrite Eqs.~(\ref{maineq11},\ref{eqn-short12}) in the form
\begin{eqnarray}
{\hat {\cal H}} \cdot \vec{\phi}_\infty &=& \alpha \sum_{l=1}^\infty
\left( \frac{(1- \delta_l^N)}{\left|l-N\right|}  -2 h(N-1) \delta_l^N \right) \phi_l.
\label{BFKLHphi}
\end{eqnarray}
To show that the $N \to \infty$ limit of this equation does reproduce the continuum BFKL kernel it is useful to work with the 
representation used in Eq.~(\ref{Mellin}), {\it i.e.},
\begin{eqnarray}
\chi_N(\gamma) &=& \sum_{l=1}^\infty
\left( \frac{(1- \delta_l^N)}{\left|l-N\right|}  -2 h(N-1) \delta_l^N \right) \left(\frac{N}{l}\right)^\gamma \nonumber\\
&=&  \left(\sum_{l=1}^{N-1}-\sum_{l=N+1}^\infty \right)
\frac{1}{N-l} \left(\frac{N}{l}\right)^{\gamma} - \sum_{l=1}^{N-1}\frac{2}{l}.
\end{eqnarray}
In Fig.~\ref{Kernels} it is numerically shown that  $\lim_{N \to \infty}\chi_N (\gamma) = \chi(\gamma)$ 
of Eq.~(\ref{chi}) in the range  $0 \leq \gamma \leq 1$. The convergence in $N$ 
is not uniform in this region since it is much faster for small values of $\gamma$.
Analytically, the continuum $N \to \infty$ limit can be found using $N = 1/ \Delta$ and $l = x / \Delta$ with 
$\Delta \to 0$ to obtain
\begin{eqnarray}
\chi_\infty  (\gamma) &=&  \int_0^1 \frac{dx}{1-x} \left(x^{-\gamma}+x^{\gamma-1}-2\right) ~=~ \chi(\gamma).
\label{wrapping}
\end{eqnarray}

\begin{figure}[htbp]
\centering
\vspace{-.5cm}
\includegraphics[width=8cm,angle=0]{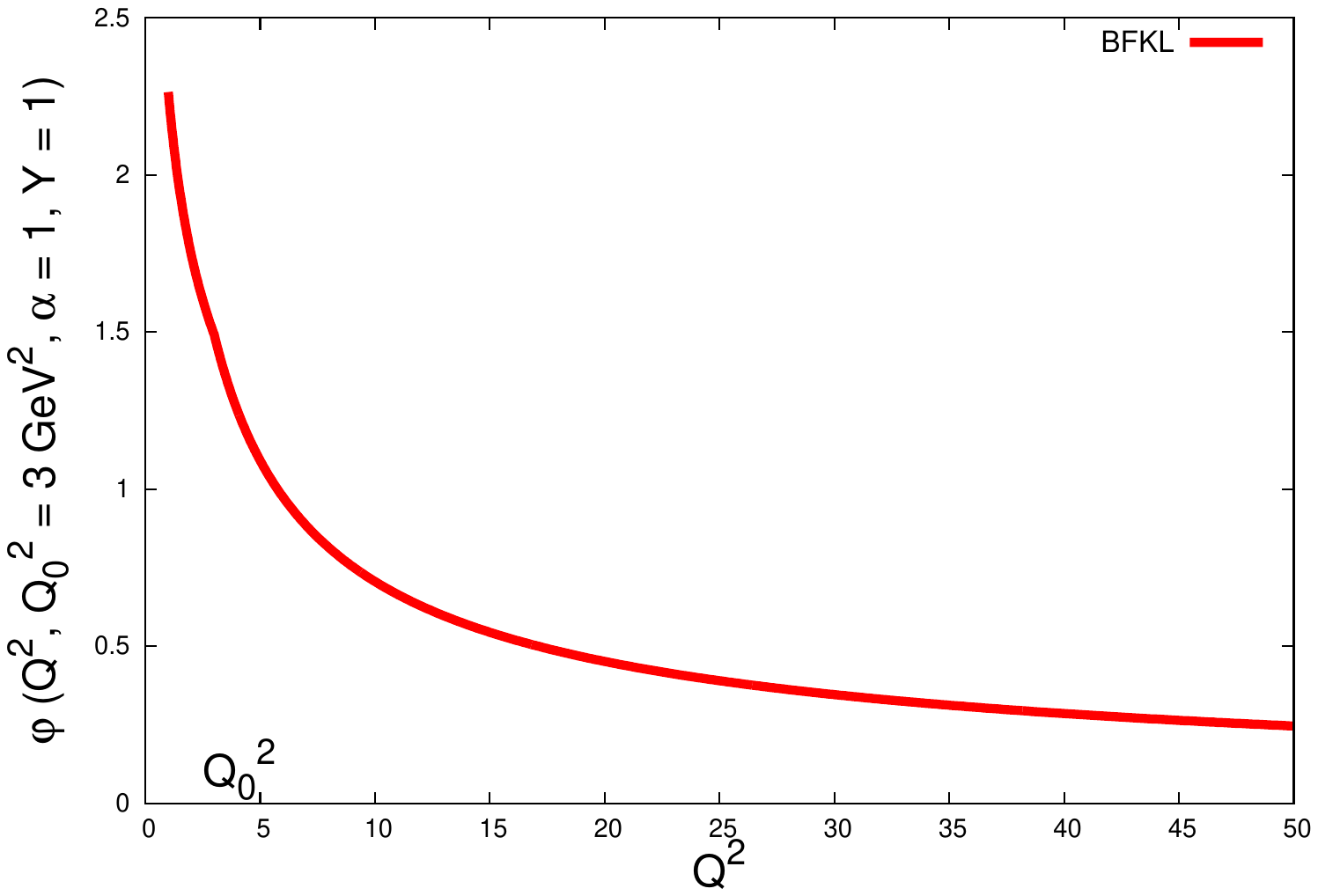}
\vspace{-1.cm}
\caption{(Anti-)Collinear behaviour of the gluon Green's function in the BFKL equation.}
\label{BFKLGGFcollinear}
\end{figure}

\begin{figure}[htbp]
\centering
\vspace{-.5cm}
\includegraphics[width=8cm,angle=0]{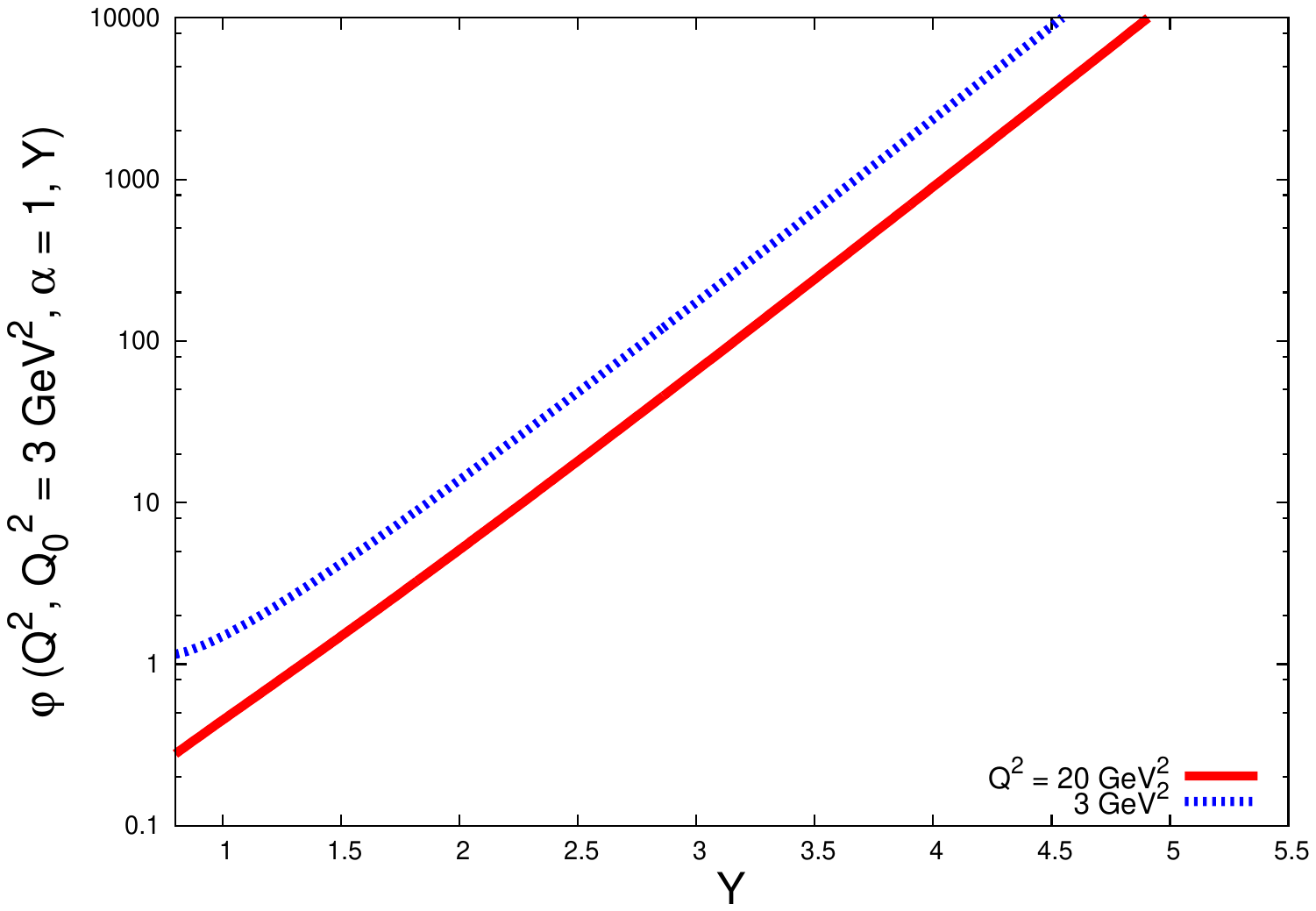}
\vspace{-1.cm}
\caption{The growth of the BFKL Green's function with $Y$.}
\label{PhigrowthYbfkl}
\end{figure}

Let us finish this section by showing the behaviour of the gluon Green's function in $Q^2$ and $Y$ space as 
obtained from Eq.~(\ref{Mellin}). In Fig.~\ref{BFKLGGFcollinear} the values  $\alpha=Y=1$ and $Q_0^2 = 3 \, {\rm GeV}^2$ have been taken,
 and the Green's function for different regions in $Q^2$, above and below the chosen $Q_0^2$, has been plotted. 
In Fig.~\ref{PhigrowthYbfkl} $Q^2$ is fixed at two different values and the growth with $Y$ for $\alpha=1$ has been shown. 
These plots will be useful for comparison with the results in Section~\ref{squareBFKL} . 

\section{Asymptotic eigensystem in a square truncation}
\label{squareBFKL}

For the exponentiation of the Hamiltonian and its action on a given initial condition state it is needed to work with a square matrix. For this the square truncation of the BFKL matrix of the form
\begin{eqnarray}
\left(\hat{\cal H}^{\rm square}\right)_{i,j} &=& 
\sum_{n=1}^{N-1} \frac{\delta_i^{j+n}}{n} +
\sum_{n=1}^{N-1} \frac{\delta_{i+n}^j}{n} - 2 h(i-1) \delta_i^j 
\end{eqnarray}
has been used for which it is possible to calculate the following vector (with $\vec{\varphi}_0 \equiv 
(\varphi^0_1,\varphi^0_2, \dots, \varphi^0_N)^t$ ):
\begin{eqnarray}
\vec{\phi} &=& e^{\alpha Y \hat{\cal H}^{\rm square}} \cdot \vec{\varphi}_0 \nonumber\\
&=& \left\{1
+ \int_0^Y dy_1 \left( \alpha \hat{\cal H}^{\rm square} \right)
+ \int_0^Y dy_1 \left( \alpha \hat{\cal H}^{\rm square} \right)
           \int_0^{y_1} d y_2 \left( \alpha \hat{\cal H}^{\rm square} \right)
\right. \nonumber\\
&+& \left.
\int_0^Y dy_1 \left( \alpha \hat{\cal H}^{\rm square} \right)
           \int_0^{y_1} d y_2 \left( \alpha \hat{\cal H}^{\rm square} \right)
           \int_0^{y_2} d y_3 \left( \alpha \hat{\cal H}^{\rm square} \right)+ \cdots
\right\} \cdot \vec{\varphi}_0,
\label{vectorphixxx}
\end{eqnarray}
which corresponds to the solution of the equation
\begin{eqnarray}
\frac{\partial \vec{\phi}}{\alpha \partial Y} &=& {\hat {\cal H}}^{\rm square} \cdot \vec{\phi}.
\end{eqnarray}
Note that one can act at both sides on 
$\vec{\phi} \equiv (\phi_1,\phi_2, \dots, \phi_N)^t$, this is different to Eq.~(\ref{eqn-short12}). 
More explicitly, in components, one can write 
\begin{eqnarray}
\frac{\partial \phi_j}{\alpha \partial Y} &=& \sum_{l=1}^{N}
\left( \frac{(1- \delta_l^j)}{\left|l-j\right|}  -2 h(j-1) \delta_l^j \right) \phi_l.
\label{compbfkl}
\end{eqnarray}

In the BFKL context each action of the Hamiltonian corresponds to a single gluon emission together with the creation of a Reggeized gluon in the $t$-channel which generates a gap in rapidity before having the next emission. In this way, powers of the Hamiltonian correspond to an increase in the gluon multiplicity. As one increases the product $\alpha Y$ more terms in the sum~(\ref{vectorphixxx}) are needed to reach convergence. 
In order to study how this picture is realised when constructing the vector $\vec{\phi}$ in the truncated BFKL case,  in the present work the  
expression in  Eq.~(\ref{vectorphixxx}) for a matrix of size $N=50$ with an initial condition $N_0=3$ (with $\varphi_i^0= \delta_i^{N_0} $) and 
a coupling $\alpha=1$ has been calculated. 
\begin{figure}[htbp]
\centering
\vspace{-.5cm}
\includegraphics[width=8cm,angle=0]{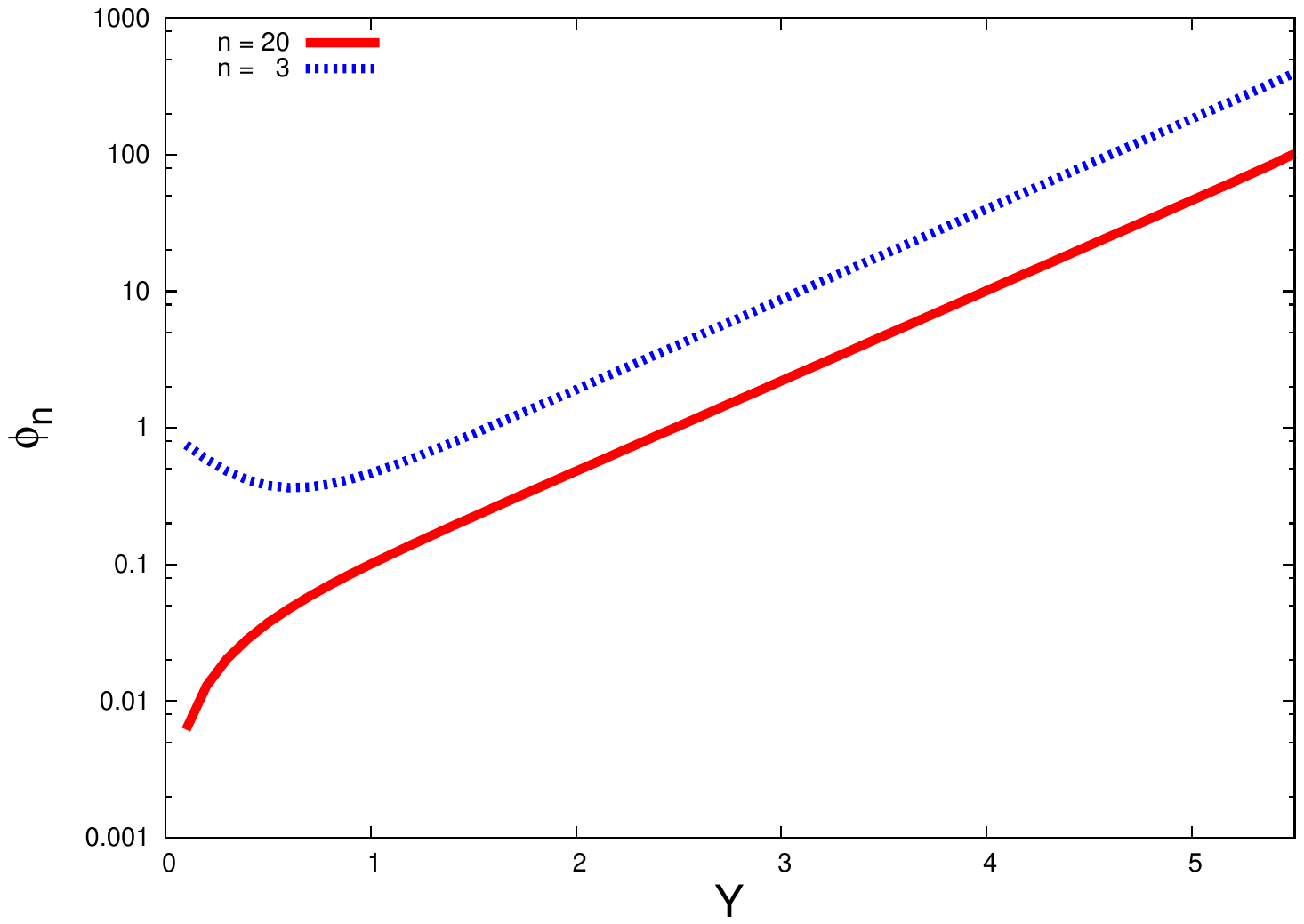}
\vspace{-1.cm}
\caption{The growth of the vector $\vec{\phi}$ with $Y$ for a matrix size $N=50$, vector components with $n=3, 20$, and $\alpha=1$.}
\label{PhigrowthYxxx}
\end{figure}
Then, a look at the $n = 3, 20$ components of the resulting $\vec{\phi}$ and the study of its dependence with $Y$ is provided in 
Fig.~\ref{PhigrowthYxxx}. It can be seen that the behaviour is very similar to that of the BFKL gluon Green's function 
in Fig.~\ref{PhigrowthYbfkl} but with a smaller growth in the truncated case (note that in BFKL the asymptotic 
growth corresponds to the Pomeron intercept 4 $\log{(2)}$).

It is possible to investigate how the asymptotic growth in the square truncation changes with the matrix size. Let us denote by 
$\vec{\psi}_{L}^{(N)}$ the $N$ eigenvectors of the $N \times N$ matrix $\hat{\cal H}^{\rm square}$, and by $\lambda_L^{(N)}$, the corresponding eigenvalues. Since any initial condition vector can be expanded in the form $\vec{\phi}_0 = \sum_{L=1}^N c_L^{(N)} \psi_L^{(N)}$, one can then write 
\begin{eqnarray}
\vec{\phi} &=& e^{\alpha Y \hat{\cal H}^{\rm square}} \cdot \vec{\phi}_0 ~=~ 
\sum_{L=1}^N c_L^{(N)} e^{\alpha Y \lambda_L^{(N)}}  \psi_L^{(N)}.
\end{eqnarray}
\begin{figure}[htbp]
\vspace{-1.cm}
\hspace{-1cm}
\includegraphics[width=15cm,angle=0]{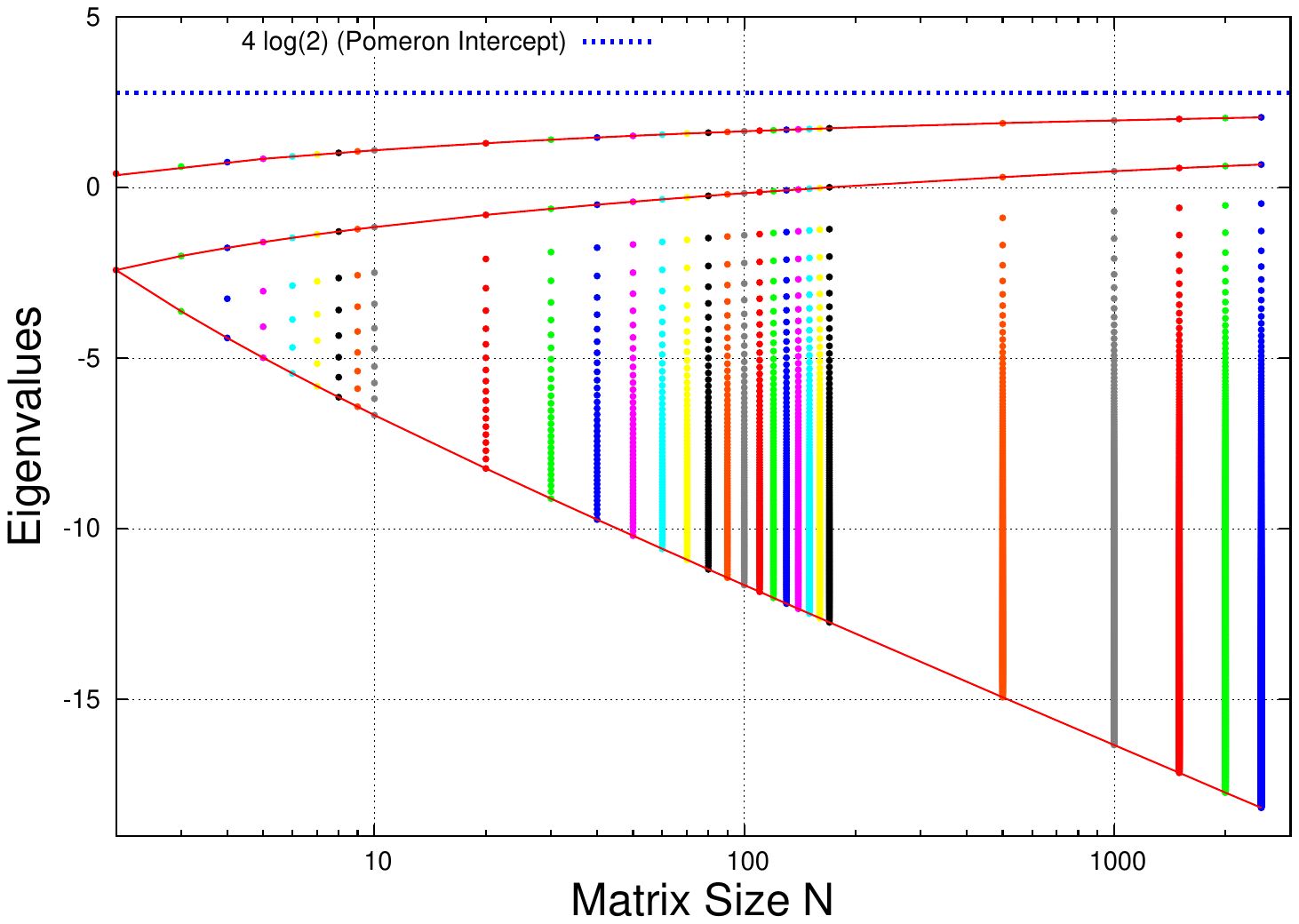}
\vspace{-1.5cm}
\caption{Dependence of the eigenvalues on the matrix size $N$.}
\label{Eigenvalues}
\end{figure}

\begin{figure}[htbp]
\vspace{-1.cm}
\begin{center}
\includegraphics[width=9cm,angle=0]{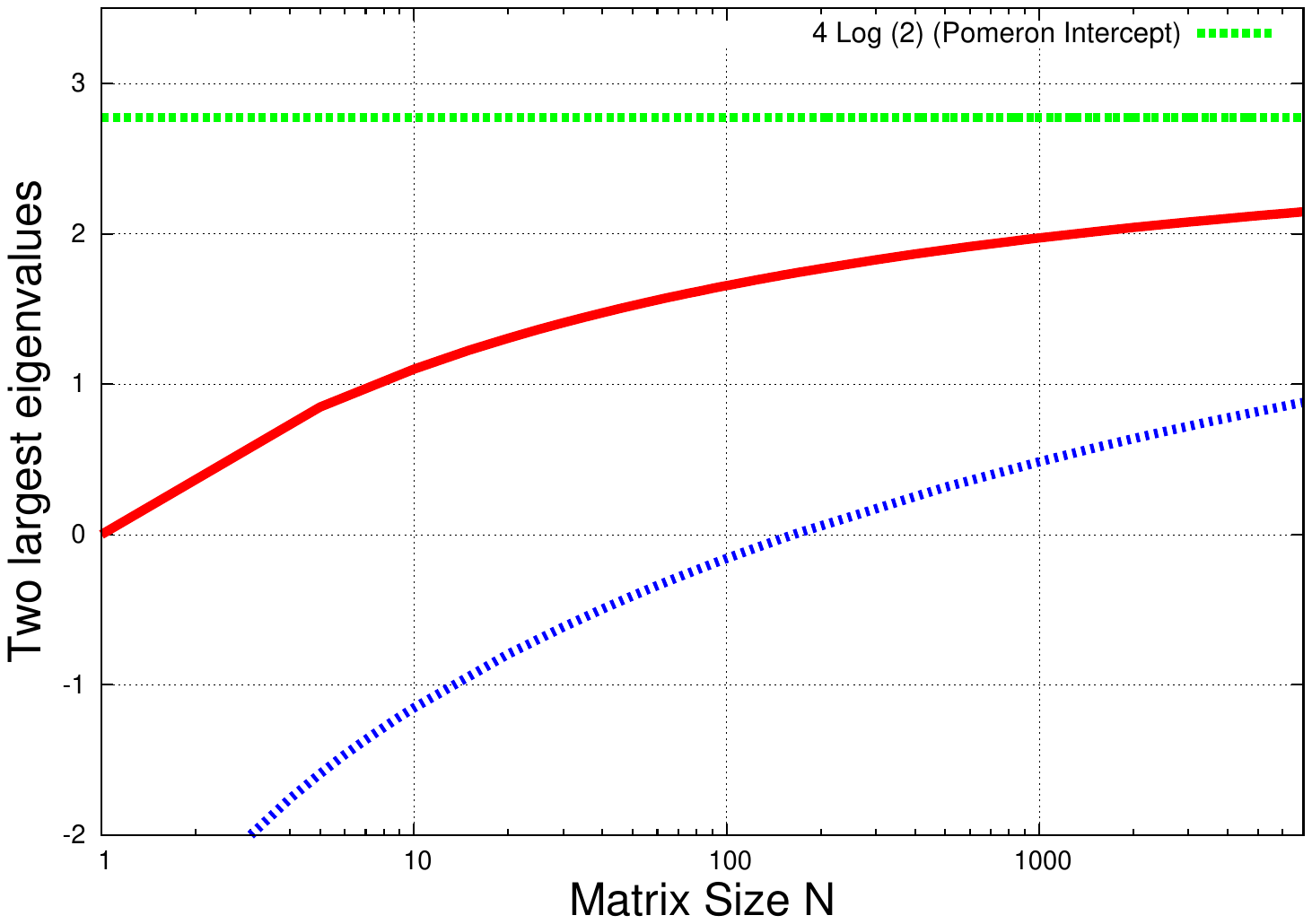}
\end{center}
\vspace{-1.cm}
\caption{Dependence of the two largest eigenvalues on the matrix size $N$.}
\label{MaxEigen}
\end{figure}

\begin{figure}[h]
\centering
\vspace{-.6cm}
\includegraphics[width=8cm,angle=0]{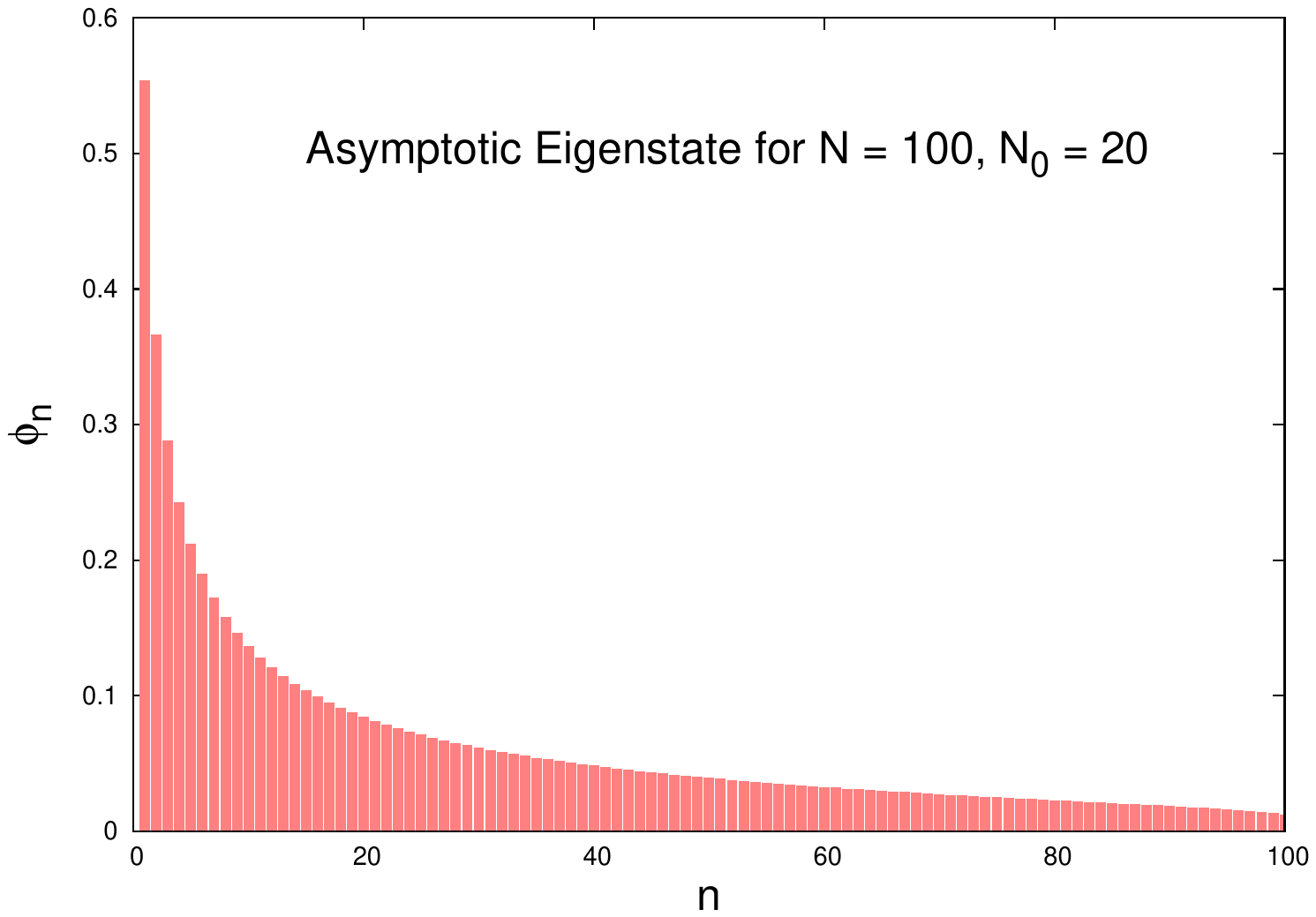}
\vspace{-1.cm}
\caption{The dominant eigenstate in the asymptotic $\alpha Y \to \infty$ region.}
\label{DominantEigenstate}
\end{figure}

The spectrum of eigenvalues of the square matrix $\hat{\cal H}^{\rm square}$ is shown in Fig.~\ref{Eigenvalues}. It can be observed that there is always a largest positive eigenvalue, $\lambda_{\rm as}^{(N)}$, 
with a gap with respect to the next one (see Fig.~\ref{MaxEigen}). 
This gap is not present for the lowest eigenvalues, which decrease 
as $\sim -2\log{(N-2)}$ for large matrix sizes.  Being the largest eigenvalue, 
$\lambda_{\rm as}^{(N)}$ drives the $\alpha Y \to \infty$ asymptotics: 
\begin{eqnarray}
\lim_{\alpha Y \to \infty} \vec{\phi} &=& c_{\rm as}^{(N)} e^{\alpha Y \lambda_{\rm as}^{(N)}}  \psi_{\rm as}^{(N)}.
\end{eqnarray}
It is interesting to note that $\lambda_{\rm as}^{(N)}$ grows very 
slowly with $N$ in a way consistent with having its $N \to \infty$ limit at the Pomeron intercept $4 \log{(2)}$ 
 (this has been confirmed up to N = 200000 where (N)
the largest eigenvalue is 2.34). We should stress at this point that one should not
expect that the square truncation of the BFKL matrix is the means of achieving
a fast convergence towards the value 4 log(2). An example of a different approach with fast numerical convergence can be found in Ref.~\cite{Hancock:1992xh}. In the present context, Fig. 7 gives a proof of concept that it is  consistent to study the eigenvalues of the square truncated BFKL matrix.  If the asymptotic eigenstate $\psi_{\rm as}^{(N)}$ is investigated one will find that the distribution of its components 
(see Fig.~\ref{DominantEigenstate}) is similar to that 
found for BFKL in Fig.~\ref{BFKLGGFcollinear}. In this analysis $N=100$ has been used, $N_0=20$ and the coupling is $\alpha=1$, but the features 
here discussed are generic. From the asymptotic expression in Eq.~(\ref{asymptotic}) one can extract the following logarithmic derivative
\begin{eqnarray}
-2 Q^2 \frac{\partial}{\partial Q^2} \varphi \left(Q^2, Y\right) &\simeq& 
1+ \frac{1}{14 \alpha Y \zeta(3)} \log{\left(\frac{Q^2}{Q_0^2}\right)}
\end{eqnarray}
It has been checked that the dominant eigenstate in the asymptotic limit 
$\alpha Y \to \infty$ follows this behaviour with a logarithmic derivative going to one as the matrix size $N$ increases. Numerically, it is much more complicated to capture the subasymptotic corrections since  
it is likely that a continuous of eigenstates contribute to them. 

\begin{figure}[htbp]
\vspace{-1.cm}
\centering
\includegraphics[width=8cm,angle=0]{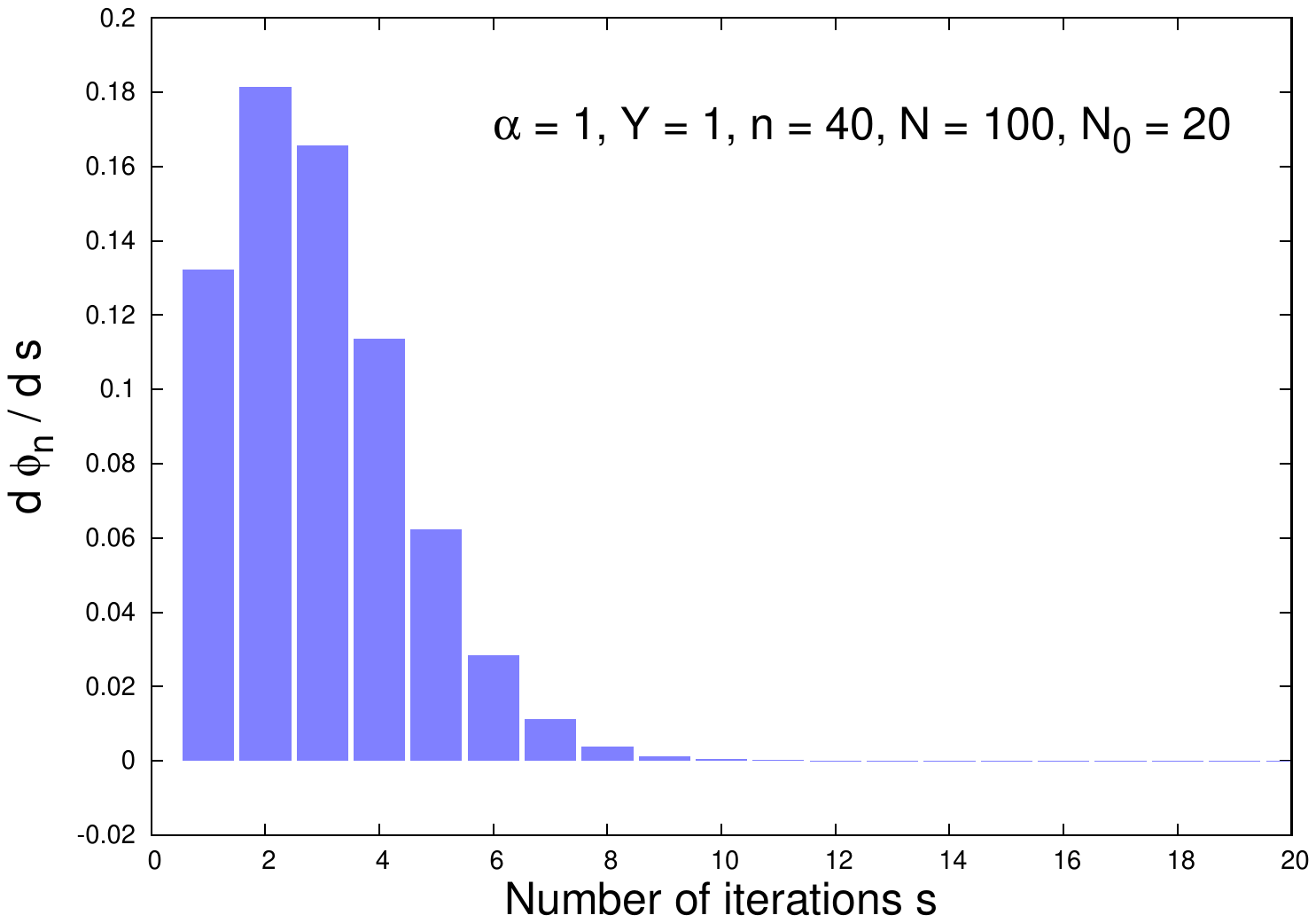}\\
\includegraphics[width=8cm,angle=0]{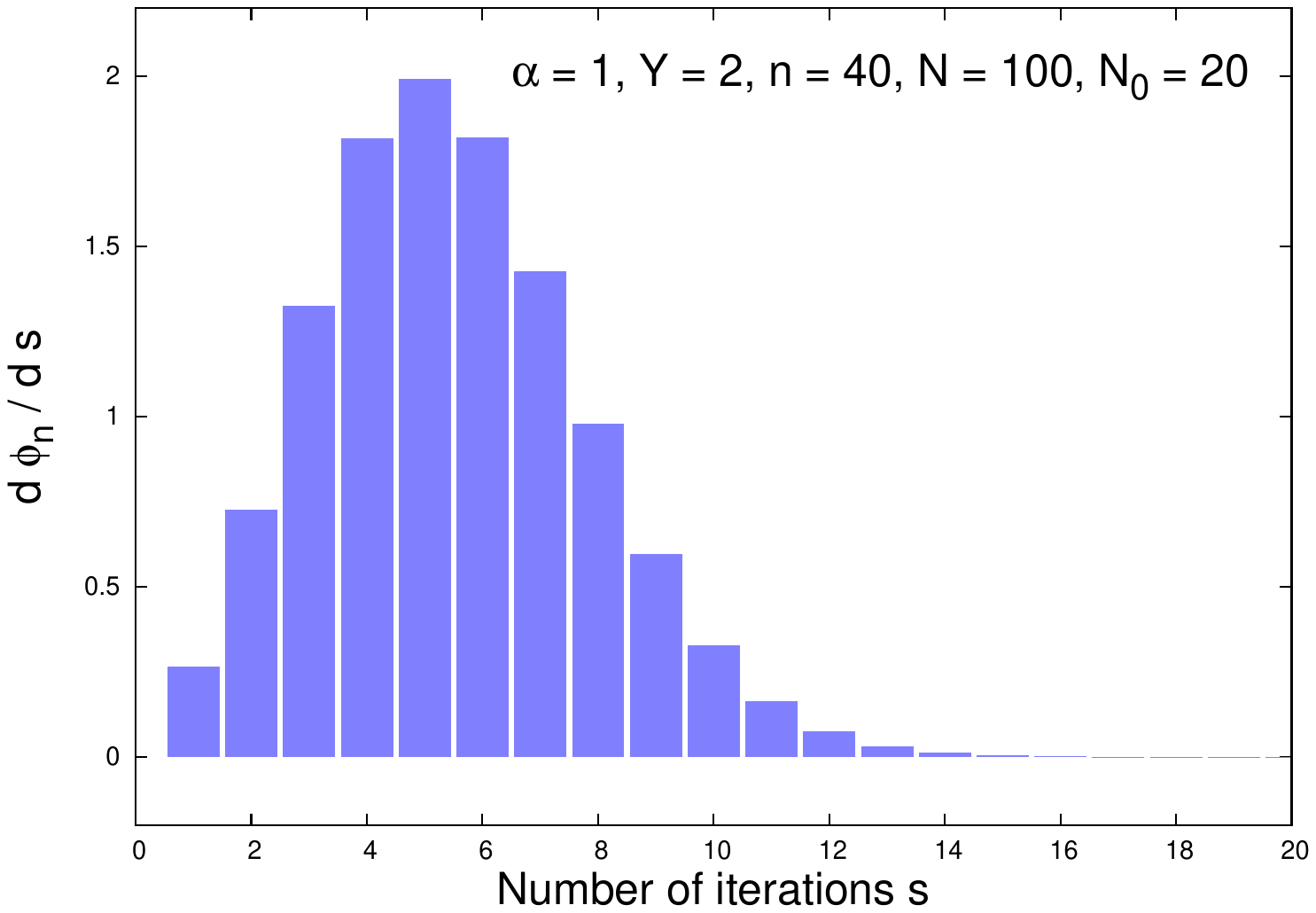}\\
\includegraphics[width=8cm,angle=0]{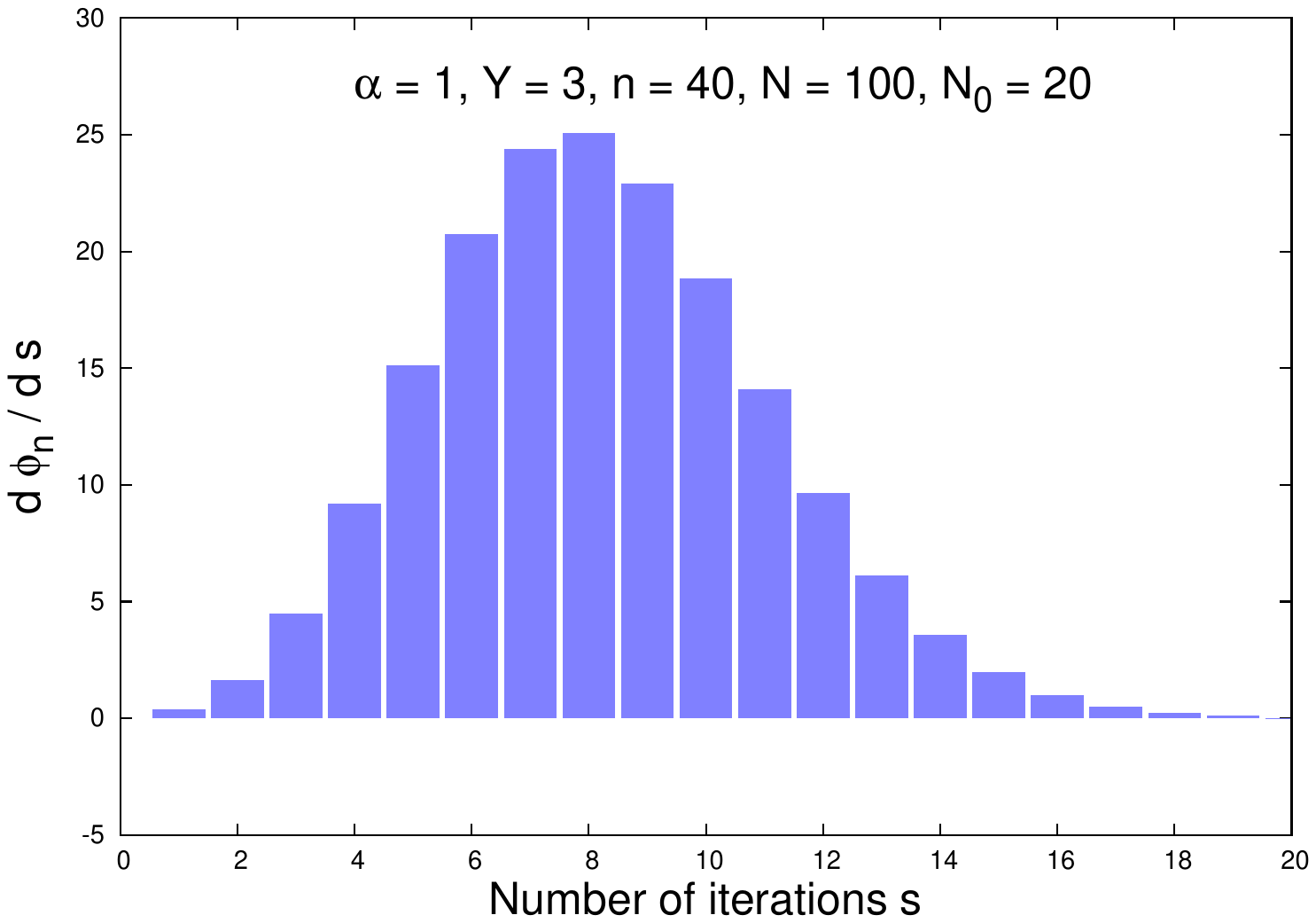}\\
\caption{Distribution in the number of iterations of the Hamiltonian for fixed ${\alpha}, N, n, Y$.}
\label{Multiplicity}
\end{figure}

It is possible to study the multiplicity distribution (average number of iterations of the Hamiltonian for a given value of $\alpha Y$ needed to reach convergence, in the QCD context this corresponds to the average number of emitted mini-jets for a fixed center-of-mass energy) in the asymptotic state 
by looking at the relative weight of the $s$ different terms in the expansion 
$e^{\alpha Y \lambda_{\rm as}^{(N)}} = \sum_{s=0}^\infty (\alpha Y \lambda_{\rm as}^{(N)})^s/s!$. This is shown for the $n=40$ component (the results are independent of this choice) of the vector $\vec{\phi}$ at different values of the rapidity variable $Y=1,2,3$ in Fig.~\ref{Multiplicity}. A Poissonian-like distribution broadening is found with an increasing mean value as $Y$ increases. These results should be 
compared to the similar ones (up to normalization) which are well-known to underly the BFKL dynamics 
as one increases $Y$ (compare with, {\it e.g.}, Fig.~1 in Ref.~\cite{Chachamis:2011nz}).

\section{Taming the energy growth by tampering with the infrared}

\begin{figure}[h]
\begin{center}
\includegraphics[width=8cm]{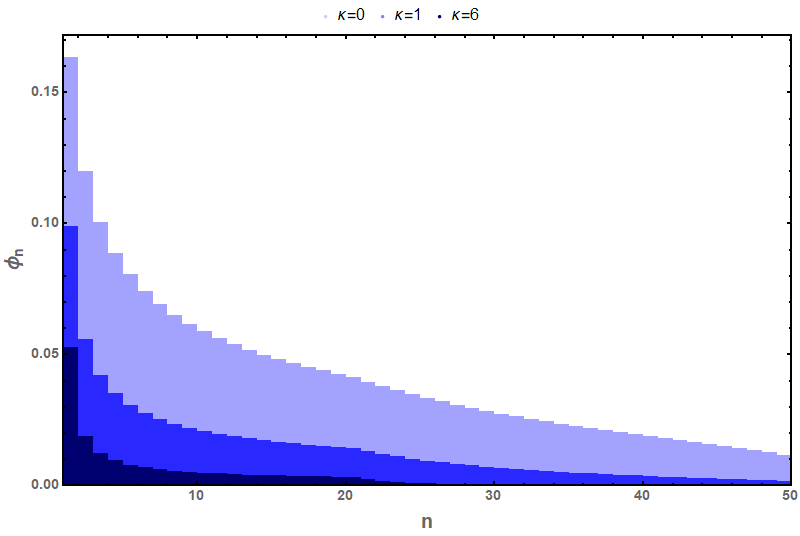}
\caption{Occupation number comparison for $\kappa=0,1,6$. $N_0=20$, $N=50$, $\Delta=1$.}
\label{ExpH}
\end{center}
\end{figure}

\begin{figure}[h]
\begin{center}
\includegraphics[width=12cm]{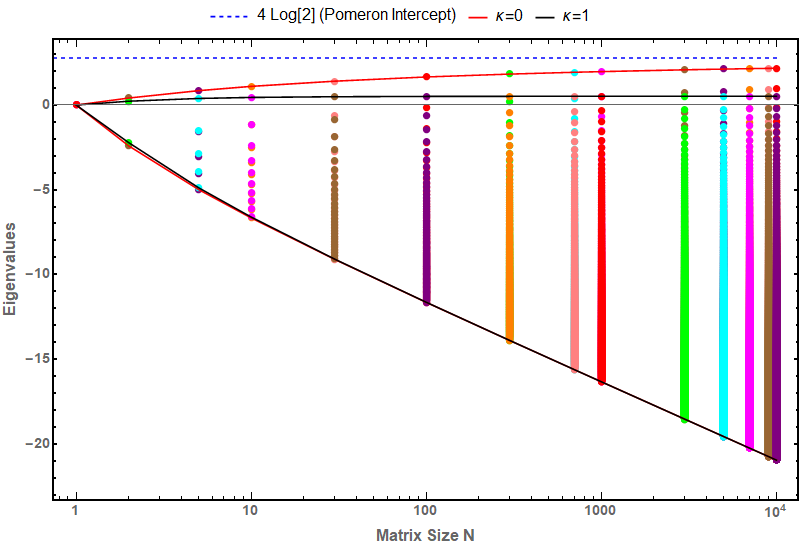}
\caption{Eigenvalues comparison for $\kappa=0$ and $\kappa=1$. N goes up to 13000.}
\label{BFKLEigenvaluesF1}
\end{center}
\end{figure}

\begin{figure}[h]
\begin{center}
\includegraphics[width=8cm]{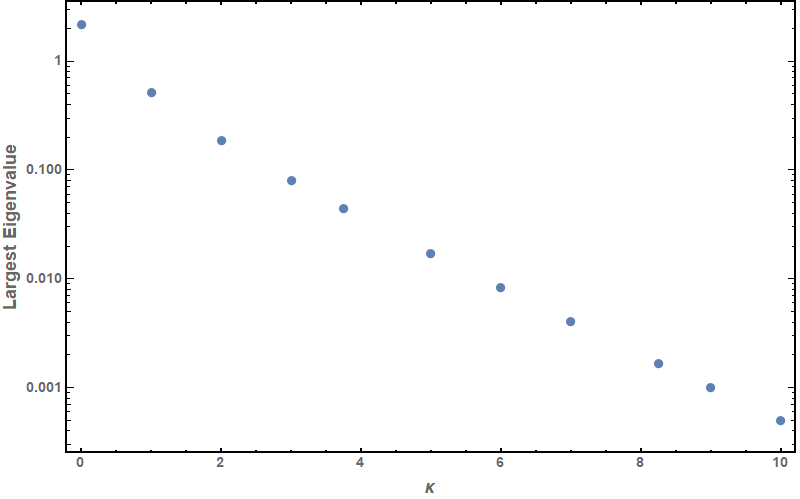}
\caption{Largest eigenvalue of the Hamiltonian for a matrix size of $N=13000$ and several values of $\kappa$.}
\label{LargestEVALNLarge}
\end{center}
\end{figure}

\begin{figure}[h]
\begin{center}
\includegraphics[width=12cm]{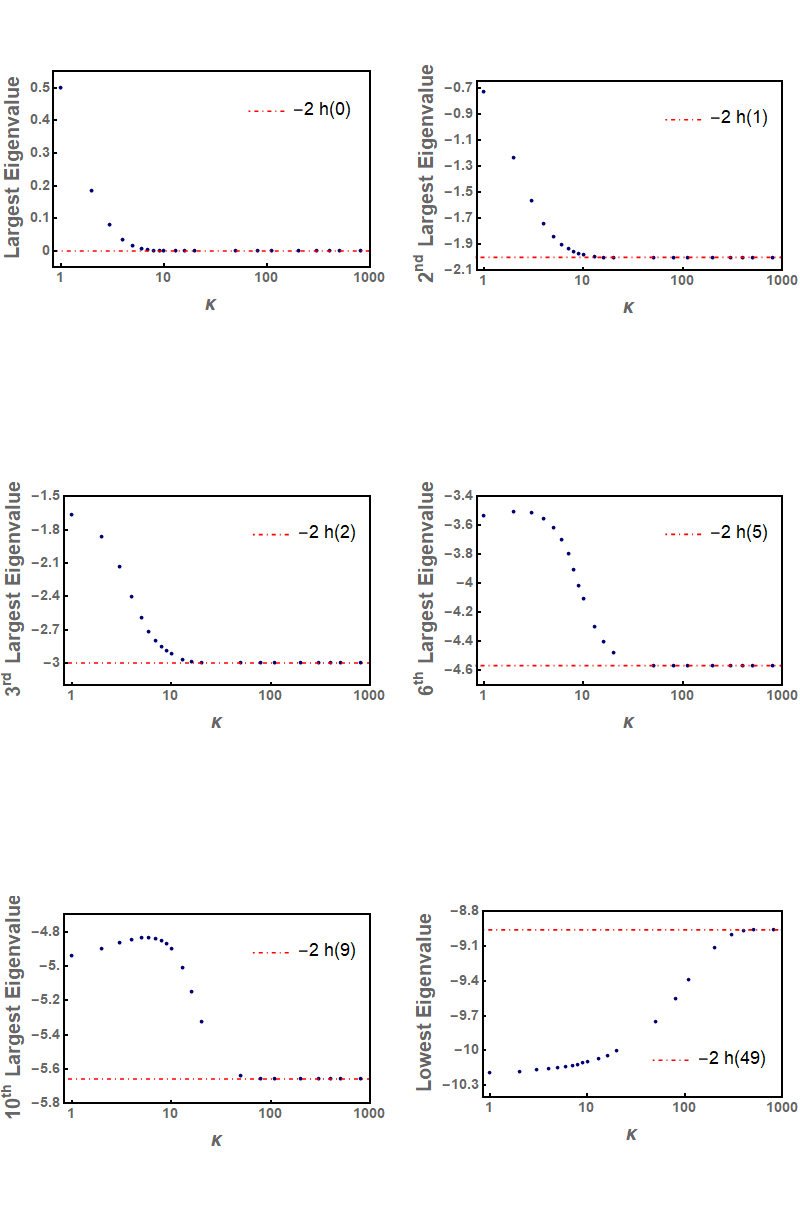}
\caption{Spectrum for a matrix size of N=50 as $\kappa$ increases.}
\label{EigenvaluesF}
\end{center}
\end{figure}

It is possible to introduce modifications in the matrix representation here presented. In particular, it is tempting to impose an absorptive constraint (see for example~\cite{Mueller:2002zm}) in the sector of the kernel responsible for the evolution into infrared modes. A precise proposal reads as follows
\begin{eqnarray}
{\partial \varphi (Q^2,Y) \over \alpha \partial Y } &=&  
\int_0^\infty {d q^2 \over |q^2-Q^2| } \Bigg[
\left(\left(\frac{q}{Q}\right)^{2\kappa} \theta(Q-q)+\theta(q-Q)\right)\varphi (q^2,Y) \nonumber\\
&-& 
{2 \, {\rm min} (q^2,Q^2) \over q^2 + Q^2 } \varphi(Q^2,Y)  \Bigg] \, .
\label{eqn-q}
\end{eqnarray}
For $\kappa = 0$ the original symmetric evolution is recovered while 
for $\kappa > 0$ a suppression of the diffusion into the infrared is imposed.  This equation can be written in the form
\begin{eqnarray}
{\partial \varphi (Q^2,Y) \over \alpha \partial Y } &=&
\int_0^1 {d x \over 1-x } \left(x^\kappa  \varphi (x \, Q^2,Y) + \frac{1}{x} 
\varphi \left( \frac{Q^2}{x},Y\right)- 2 \, \varphi(Q^2,Y)  \right) ,
\end{eqnarray}
with solution in the region $0 < \Re (\gamma) < 1$
\begin{eqnarray}
\varphi (Q^2,Y) &=& \int_{a - i \infty}^{a+i \infty} \frac{d \gamma}{ 2 \pi i} 
\left({Q^2 \over Q_0^2}\right)^{\gamma-1} e^{\alpha Y \chi_\kappa (\gamma)}, \\
\chi_\kappa (\gamma) &=&  2 \psi(1) - \psi(\gamma+\kappa) - \psi (1-\gamma).
\end{eqnarray}
The discretised representation can be written as
\begin{eqnarray}
\frac{\partial \phi_N}{\alpha \partial Y} &=&  \sum_{n=1}^{N-1} \left(\frac{n}{N}\right)^\kappa
\frac{\phi_n }{N -n }+\sum_{n=N+1}^\infty \frac{\phi_n }{n-N }- 2 h(N-1)  \phi_N,
\label{maineq}
\end{eqnarray}
with the corresponding matrix Hamiltonian being 
\begin{equation}
\hat{\mathcal{H}}_\kappa=
\begin{pmatrix}
-2h(0) & 1 & \frac{1}{2} & \frac{1}{3} & \dots \\
1 ( \frac{1}{2})^\kappa & -2h(1) & 1 & \frac{1}{2} & \dots \\
\frac{1}{2} ( \frac{1}{3})^\kappa &  1  ( \frac{2}{3})^\kappa & -2h(2) & 1 & \dots  \\
 \frac{1}{3}  ( \frac{1}{4})^\kappa & \frac{1}{2}  ( \frac{2}{4})^\kappa & 1  ( \frac{3}{4})^\kappa & -2h(3) & \dots \\
\vdots & \vdots & \vdots & \vdots & \vdots \\
\frac{1}{N-1} ( \frac{1}{N})^\kappa & \frac{1}{N-2}  ( \frac{2}{N})^\kappa & \frac{1}{N-3}  ( \frac{3}{N})^\kappa & \frac{1}{N-4}  ( \frac{4}{N})^\kappa & \dots
\end{pmatrix}.
\end{equation}
The associated matrix elements are
\begin{eqnarray}
\Big ( \hat{\mathcal{H}}_\kappa \Big )_{i,j} &=& \sum_{n=1}^{N-1} \frac{\delta_i^{j+n}}{n} \Big ( \frac{j}{i} \Big )^\kappa + \sum_{n=N+1}^{\infty} \frac{\delta_{i+n}^j}{n} - 2h(i-1)\delta_i^j \nonumber\\
&=&\frac{1}{i-j} \Big ( \frac{j}{i} \Big )^\kappa \theta(i-j) + \frac{1}{j-i} \theta(j-i) -2h(i-1) \delta_i^j \, .
\label{components01}
\end{eqnarray}
It is now instructive to study the action of its square truncation on an initial condition vector with $N_0=20$ and, again, $N=50$, 
$\Delta=1$. This is plotted in Fig.~\ref{ExpH} for values of the 
new parameter $\kappa=0,1,6$.
When $\kappa=0$ the usual BFKL behaviour for the Green's function is obtained. As $\kappa$ increases its value is strongly suppressed due to the infrared barrier introduced in the evolution kernel. As previously discussed, the asymptotic behaviour is governed by the spectrum of eigenvalues of the matrix Hamiltonian. This is shown in 
Fig.~\ref{BFKLEigenvaluesF1}  where it can be seen how introducing 
$\kappa=1$ makes the asymptotic eigenvalues decrease drastically. This includes the case of the largest eigenvalue whose dependence 
on $\kappa$ is plotted, for a large matrix size of $N=13000$, in 
Fig.~\ref{LargestEVALNLarge}.  Since the largest eigenvalue rapidly tends to zero as $\kappa$ increases, this limit can be considered as an effective method to generate an evolution with a weaker violation of the Froissart bound. 
There is an interesting interpretation of the spectrum shown in 
Fig.~\ref{BFKLEigenvaluesF1} when $\kappa \to \infty$. In this limit 
the Hamiltonian becomes upper triangular and its eigenvalues correspond to the diagonal matrix elements, $-2 h(i-1)$, with $1 \leq i \leq N$. To illustrate this point, the behaviour of some of the eigenvalues as $\kappa$ grows is given in Fig.~\ref{EigenvaluesF}. 

\begin{figure}[h]
\begin{center}
\includegraphics[width=12cm]{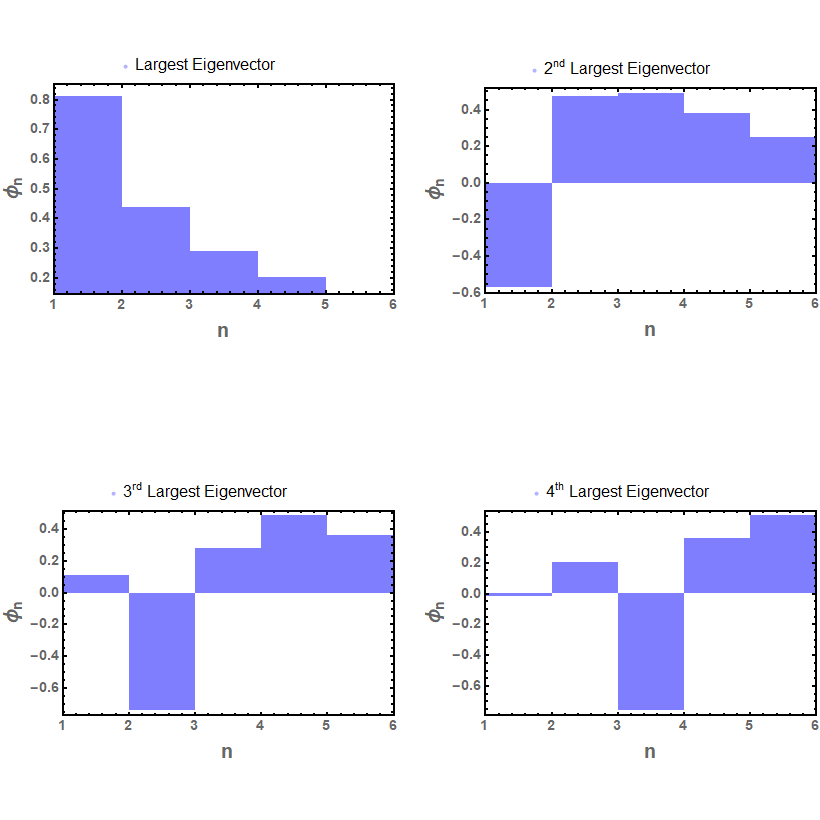}
\caption{Eigenvector coefficients for $\kappa=0$, $N=5$.}
\label{EvectorF0N5}
\end{center}
\end{figure}

To conclude this section, a brief study of the associated eigenvectors  can be introduced. It is worth pointing out that the eigenvector linked to the largest eigenvalue is the only one with all its components being positive, as can be seen for a small matrix size with $N=5$ and $\kappa=0$, in Fig.~\ref{EvectorF0N5}. 
In this plot a second interesting feature appears: the secondary eigenvectors show oscillatory behaviour in their components. This implies that there exists a double mechanism to suppress the subleading eigenvalues: firstly, the fact that they are numerically smaller and, secondly, the destructive interference among their eigenvectors. These qualitative structure holds for large matrices and 
different values of $\kappa$.

\section{Matrix representation of the sl(2) spin chain Hamiltonian}

As it was mentioned in the Introduction, there has been big progress in the understanding of the planar limit of the 
${\cal N} = 4$ SYM dilatation operator by mapping it to the Hamiltonian of an integrable one-dimensional 
spin chain. The part of the theory of interest in this work is the non-compact bosonic 
sl(2) closed subsector where states are constructed with scalar fields $\Phi$ (SO(6) Yang-Mills bosons) and their spacetime covariant derivatives ${\cal D} \Phi$, which scale under the sl(2) subgroup of the Lorentz group. The corresponding charges are related to the SO(4,2) group and to the SO(6) ${\cal R}$-symmetry. 

The commutation relations of the sl(2) subalgebra of the superconformal algebra are $\left[J^{(+)},J^{(-)} \right]=
-2 J^{(3)}, \left[J^{(3)},J^{(\pm)}\right] = \pm J^{(\pm)}$. The spin chain Hamiltonian 
representing the one-loop anomalous dimensions of ${\cal N} = 4$ SYM operators with spin $S-1$ in 
the planar limit of the sl(2) sector reads
\begin{eqnarray}
{\cal H}_{1,2}^{{\rm sl}(2)}  \theta(S-N) (a_1^\dagger)^{N-1}(a_2^\dagger)^{S-N} \ket{00} &=& \nonumber\\
&&\hspace{-5.5cm}- \lambda \sum_{l=1}^\infty
\left(\frac{(1- \delta_l^N)}{\left|l-N\right|} -  \left(h (N-1) + h (S-N)\right) \delta_l^N \right)
\theta(S-l) (a_1^\dagger)^{l-1}(a_2^\dagger)^{S-l} \ket{00},
\label{BeisertH}
\end{eqnarray}
where $\lambda= \frac{g^2 N_c}{8 \pi^2}$ is the coupling. The notation in terms of harmonic oscillators $(a^\dagger)^n  \ket{0} = \frac{1}{n!} ({\cal D})^n \Phi$ 
(with $a \ket{0} = 0$, $\left[a,a^\dagger\right]=1$), which corresponds to a site in a one-dimensional lattice (the total number of lattice sites is equal to the total ${\cal R}$-charge), has been 
used. $n$ indicates that a given site is in the $n$-th excited state with respect to the Tr$(\Phi^2)$ vacuum.  
These excitations are 
classified in the spin $s=-\frac{1}{2}$ representation of sl(2), which is infinite-dimensional. 
There is an overall trace in the operator cyclically ordering the different sites in the spin chain.
Eq.~(\ref{BeisertH}) was introduced by Beisert in~\cite{Beisert:2003jj} and it corresponds to the nearest-neighbor 
one-loop Hamiltonian of an integrable XXX spin $s= - \frac{1}{2}$  chain. More explicitly, this Hamiltonian is invariant under the sl(2) generators 
\begin{eqnarray}
J^{(+)}_{12} &=& a^{\dagger}_1 (1 + a^{\dagger}_1 a_1) + a^{\dagger}_2  (1 + a^{\dagger}_2 a_2),\\ 
J^{(-)}_{12} &=& a_1 + a_2, \\ 
J^{(3)}_{12} &=& 1 + a^{\dagger}_1 a_1 + a^{\dagger}_2 a_2.  
\end{eqnarray}

There is an interesting similarity between this spin chain representation and the square truncation of the BFKL equation. In order to find it it is needed to 
focus on a ``slice" of the XXX$_{-\frac{1}{2}}$ spin chain Hamiltonian characterized by $S-N = N-1$. In this case the spin chain Hamiltonian acts on a diagonal state with the same number of derivatives in each oscillator (both are in the same $(N-1)$-th excited state):
\begin{eqnarray}
{\cal H}_{1,2}^{{\bf sl}(2)}  (a_1^+)^{N-1}(a_2^+)^{N-1} \ket{00} &=& \nonumber\\
&&\hspace{-3cm}- \lambda \sum_{l=1}^{2N-1}
\left(\frac{(1- \delta_l^N)}{\left|l-N\right|} -2 h(N-1) \delta_l^N \right) (a_1^+)^{l-1}(a_2^+)^{2N-1-l} \ket{00}.
\label{Beisertdiag}
\end{eqnarray}
Let us now compare this expression with that for the square truncation of the BFKL Hamiltonian in Eq.~(\ref{compbfkl}). It is striking that the terms under the sum in Eq.~(\ref{Beisertdiag}) and Eq.~(\ref{compbfkl}) are identical if one identifies $\phi_l$ with $(a_1^+)^{l-1}(a_2^+)^{2N-1-l} \ket{00}$ and $\alpha$ with $- \lambda$. 
It should be  stressed that in Eq.~(\ref{BeisertH}) the diagonal $h(N-1)+h(S-N)$ terms coincide with those of 
BFKL only for $S = 2 N -1$.

The main difference between the discretized-in-virtuality BFKL equation and the sl(2) spin chain projected on a diagonal state 
stems from the components with $l \geq 2 N$ in Eq.~(\ref{BFKLHphi}) which are not present in Eq.~(\ref{Beisertdiag}). The corresponding matrices read 
\begin{eqnarray}
\hat{\cal H} =
\left(\hspace{-0.cm}\begin{array}{ccccc}
-2 h(0) & 1 & \frac{1}{2} & \cdots&\frac{1}{N-1}\\
1& - 2 h(1) & 1 &\cdots&\frac{1}{N-2}\\
\frac{1}{2}&1& -2 h (2) &  \cdots&\frac{1}{N-3}\\
\vdots&\vdots&\vdots&  \ddots&\vdots\\
\frac{1}{N-1}&\frac{1}{N-2}&\cdots&  \cdots& -2 h (N-1)\\
\end{array}\hspace{-0cm}\right)_{{\rm XXX}_{-\frac{1}{2}}}
\hspace{-0.6cm}\left.\begin{array}{ccc}
\frac{1}{N} & \frac{1}{N+1} &\cdots\\
\frac{1}{N-1}& \frac{1}{N}  &\cdots\\
\frac{1}{N-2}&\frac{1}{N-1}&\cdots\\
\vdots&\vdots&\ddots\\
1&\frac{1}{2}&\cdots\\
\end{array}\right)_{\rm BFKL}
\label{twomatrices}
\end{eqnarray}
However, the square truncation of the BFKL matrix coincides with the slice of the spin 
chain under study in this work. 

\begin{figure}[h]
  \centering
  \vspace{-.5cm}
  \includegraphics[width=8cm,angle=0]{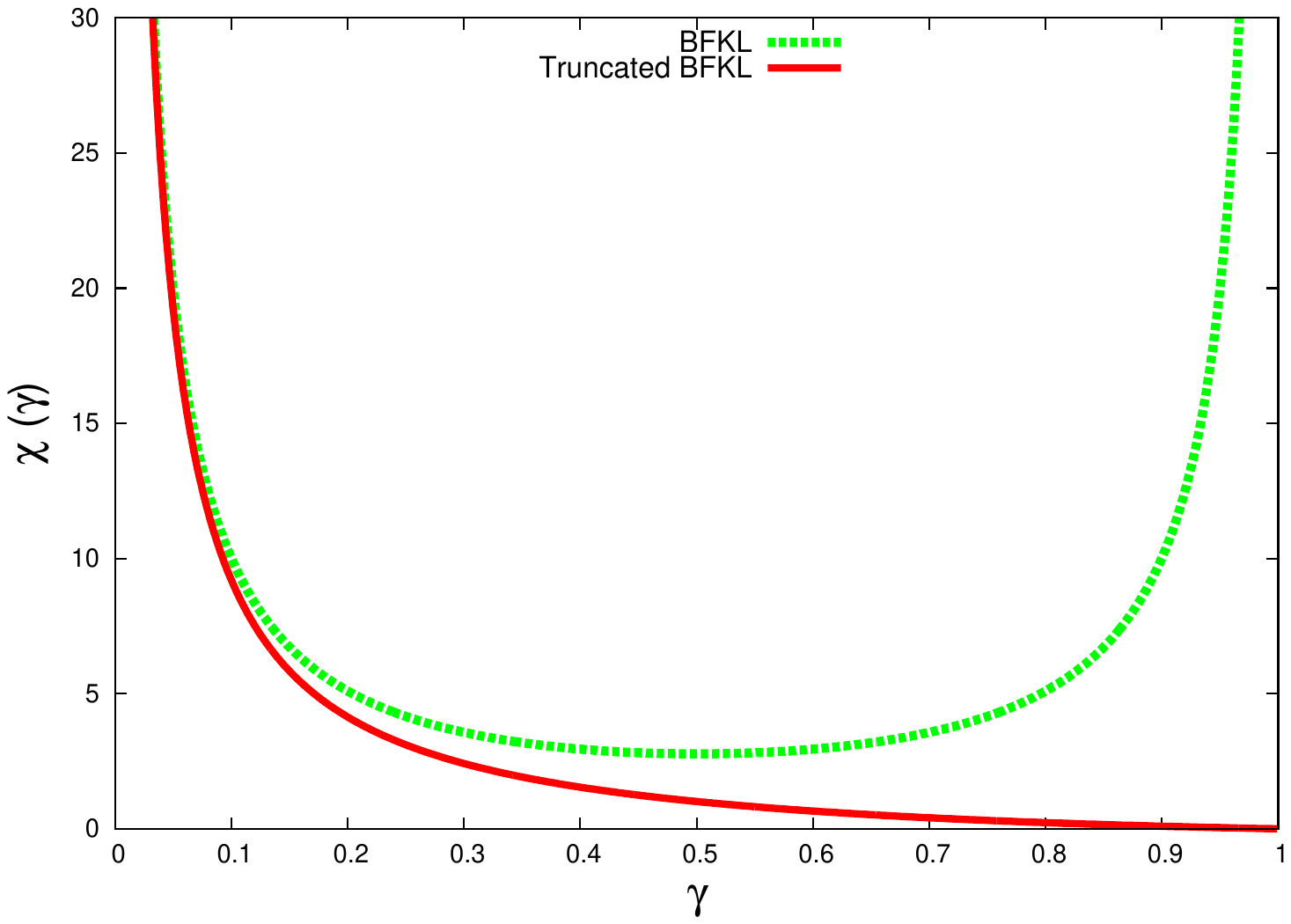}
  \vspace{-1.cm}
  \caption{The BFKL and the truncated BFKL kernels in $\gamma$ space.}
  \label{xxxvsbfkl}
\end{figure}

\begin{figure}[h]
  \centering
  \vspace{-.5cm}
  \includegraphics[width=8cm,angle=0]{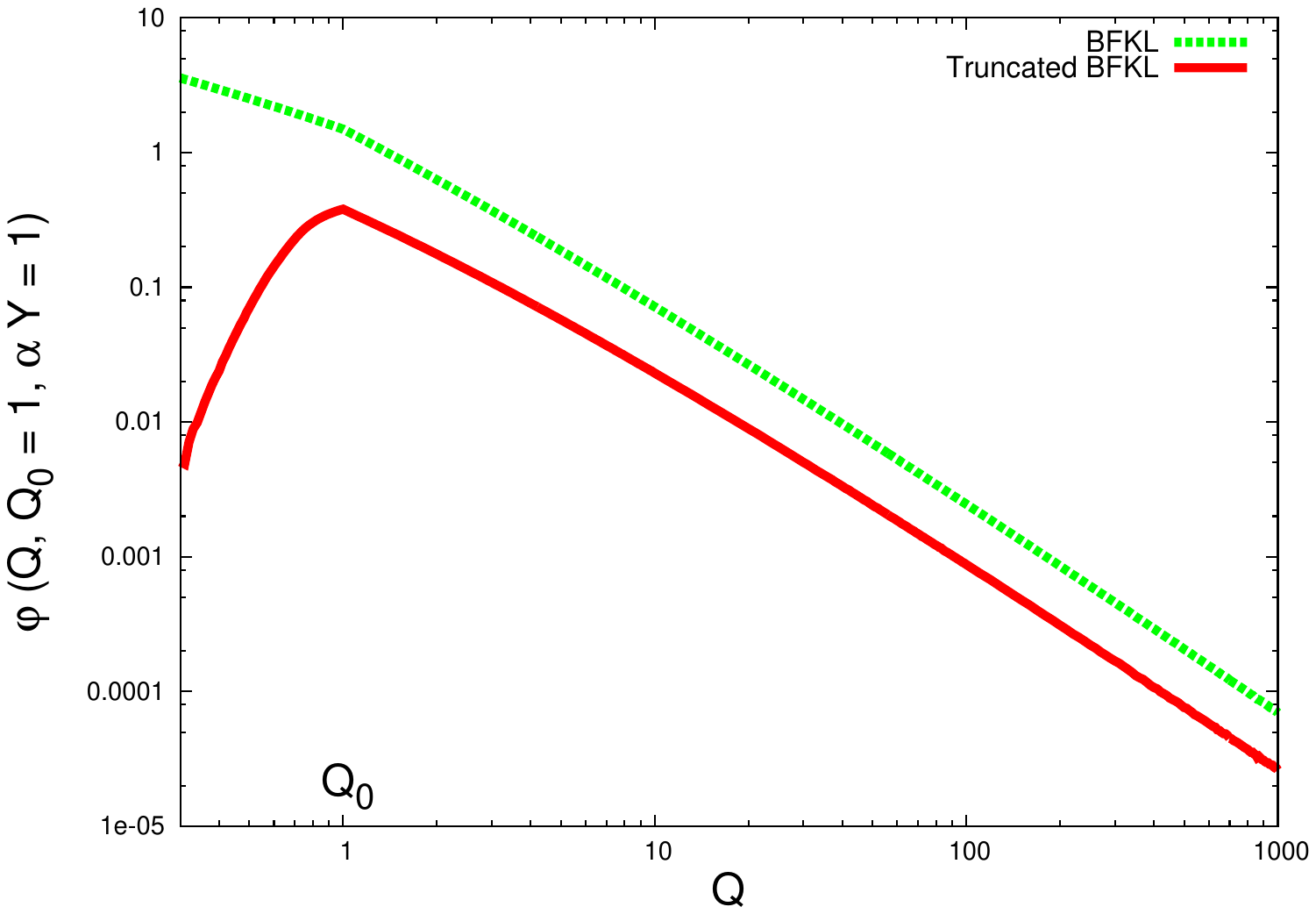}
  \vspace{-1.cm}
  \caption{Collinear behaviour of the BFKL and the truncated BFKL gluon Green's functions.}
  \label{GGFxxxvsbfkl}
\end{figure}

Let us stress that the truncation of the BFKL kernel leads to a new evolution equation with different eigenvalues  and a different Green's function. In order to make this point more clear one can investigate the corresponding evolution equation in the continuum limit. We introduce a change of variables in 
Eq.~(\ref{eqn-q5}), namely, $q^2 \rightarrow l^2+ Q^2$ and in addition we change the upper limit of the integration from infinity to $\bar{Q}^2$ 
such that we finally have:
\begin{eqnarray}
{\partial \varphi (Q^2,Y) \over \alpha \partial Y } 
&=& \int_0^1 {d x \over 1-x } \left\{\varphi (x \, Q^2,Y) + \frac{1}{x}
\varphi \left( \frac{Q^2}{x},Y\right)- 2 \, \varphi(Q^2,Y)  \right\} \nonumber\\
&-&  \int_0^{\frac{Q^2}{Q^2+{\bar Q}^2}} {d x \over (1-x) x } \varphi \left(\frac{Q^2}{x},Y\right) 
\label{QQbar}
\end{eqnarray}
where ${\bar Q}^2 \to \infty$ leads to the  usual BFKL equation and ${\bar Q}^2 = Q^2$ to its square truncation in Eq.~(\ref{compbfkl}) with $N \to \infty$.
Making use of  
the representation for $\varphi$ of Eq.~(\ref{Mellin}) in Eq.~(\ref{QQbar}) one obtains, in the $0 < \gamma < 1$ region, the following expression 
\begin{eqnarray}
\chi(\gamma) &=& \frac{1}{\gamma} +  2 \sum_{L=1}^\infty \zeta (2 L +1) 
\gamma^{2L}  + \sum_{L=1}^\infty \frac{1}{\gamma-L} \left(\frac{Q^2}{Q^2+{\bar Q}^2}\right)^{L-\gamma}. 
\end{eqnarray}
This leads to a simple relation between the BFKL and the truncated BFKL kernels:
\begin{eqnarray}
\chi^{\rm square} (\gamma) &=& \chi^{\rm BFKL} (\gamma) + \sum_{L=1}^\infty \frac{2^{\gamma-L}}{\gamma-L}.
\end{eqnarray}
The main feature of this result is that the new term in $\chi^{\rm square}$ contains a pole at $\gamma \to 1$ which 
cancels a similar one in $\chi^{\rm BFKL}$. The $\gamma \to 0$ region only receives finite corrections. This can be seen in 
Fig.~\ref{xxxvsbfkl}. 
The asymmetry in the square truncation of the kernel generates a different behaviour in the collinear $Q>Q_0$ and anti-collinear $Q<Q_0$ 
regions of the Green's function $\varphi (Q,Q_0,Y)$. This is explicitly 
shown in Fig.~\ref{GGFxxxvsbfkl} where the Green's function is shown for both equations  for $Q_0=1$ and $\alpha Y=1$.
Both solutions have a very different structure when $Q<Q_0$, or $N < N_0$ in the discretized 
version. Their behaviour is more similar in the asymptotic region $Q \gg Q_0  (N \gg N_0)$ since there they 
share  the same leading singularity at $\gamma \to 0$:
\begin{eqnarray}
\chi^{\rm BFKL} (\gamma) 
&=& \frac{1}{\gamma} +  2 \sum_{L=1}^\infty \zeta (2 L +1) \gamma^{2L},
\label{chibfkl}\\
\chi^{\rm square} (\gamma) &=& \frac{1}{\gamma} +  2 \sum_{L=1}^\infty \zeta (2 L +1) \gamma^{2L} 
- \sum_{m=0}^\infty \gamma^m \sum_{n=0}^m
\frac{(\log{2})^n}{n!} {\rm Li}_{1+m-n} \left(\frac{1}{2}\right).
\label{chixxx}
\end{eqnarray}
It is worth pointing out that these kernels generate different leading ${\cal O}\left((\alpha / \omega)^n\right)$ contributions to the anomalous 
dimension of twist-two operators with spin $M=\omega-1$, for $\omega \to 0$:
\begin{eqnarray}
\gamma^{\rm BFKL}_\omega &=& \frac{\alpha}{\omega} + 2 \zeta(3) \left(\frac{\alpha}{\omega} \right)^4 
+ 2 \zeta(5) \left(\frac{\alpha}{\omega} \right)^6 
+ 12 \zeta(3)^2 \left(\frac{\alpha}{\omega} \right)^7 +\dots, \nonumber\\
\gamma_\omega^{\rm square} &=& \frac{\alpha}{\omega} - \log{2} \left(\frac{\alpha}{\omega} \right)^2
+  \frac{1}{2} \left((\log{2})^2 - \zeta(2) \right)  \left(\frac{\alpha}{\omega} \right)^3 \nonumber\\
&+& \left(\frac{1}{3} (\log{2})^3 + \frac{3}{2} \zeta(2) \log{2}+ \frac{9}{8} \zeta(3) \right)  \left(\frac{\alpha}{\omega} \right)^4 + \cdots
\end{eqnarray}

Let us conclude with some possible connections to other works like that in~\cite{Shuvaev:2006br} where an expression as in Eq.~(\ref{eqn-q}) was found with $\varphi$ corresponding there to the distribution of soft photons in a charged source, with no relation to high energy scattering in the Regge limit or the BFKL equation. A similar analysis can be applied in that context. From a more formal  point of view, it is likely that a link can be found between the matrix representation of the 
sl(2) Hamiltonian here unveiled and the work in~\cite{Adler}, where it was 
investigated how the discrete sinh-Gordon equation leads to the Toeplitz 2-Toda lattice. This new lattice has 
$\tau$-functions which are annihilated by operators living in a SL(2,$\mathbb{Z}$) subalgebra of the Virasoro algebra, and have 
a very similar structure to Eq.~(\ref{Ston}) if the harmonic weights $1/n$ are identified with time variables in the 
IR/UV directions. It is on these time variables where the Virasoro operators act. Finally, in~\cite{Bellucci:2004qr} a coherent state representation for the Hamiltonian of the spin chain with sl(2) symmetry 
was derived. The action of the Hamiltonian on the coherent states $\vec{n}_1, \vec{n}_2$ then reads
\begin{eqnarray}
\left<\vec{n}_1, \vec{n}_2 \right| {\cal H}_{1,2}^{{\rm sl}(2)} \left| \vec{n}_1, \vec{n}_2 \right>  &=& 
\log{\left(1-\frac{(\vec{n}_1 - \vec{n}_2)^2}{4}\right)},
\end{eqnarray}
with $\vec{n}_{1,2}$ living on a two-dimensional hyperboloid for sl(2). It will be interesting to study 
the relation of this work with the results here presented.

\section{Conclusions}

In this work a semi-infinite matrix representation of the BFKL equation has been given in 
Eq.~(\ref{matrixH}), together with its physical interpretation in terms of a 
diffusion process into infrared and ultraviolet regions of virtuality space. 
It has been found that the square truncation of the semi-infinite matrix in the BFKL equation and the action of the 
$s=-\frac{1}{2}$ XXX spin chain Hamiltonian on a symmetric double copy of the harmonic oscillator share many common features. 
It was possible to make this connection by taking the original BFKL equation to its 
forward limit and averaging over the azimuthal angle dependence of its kernel. The 
remaining physical variable, which has been discretised to write the  matrix representation, corresponds to the virtuality of exchanged gluons. 
The associated gluon Green's function has been constructed using a square truncation of the BFKL matrix, by exponentiating it and acting on a general initial condition. It has been shown that in this case both systems manifest the same asymptotic behaviour. Finally, a modification of the $t$-channel gluon propagators in the infrared has been proposed as a 
simple mechanism to generate an evolution with a weaker violation of the Froissart bound by taming the exponential growth of the original evolution equation.  
$~$\\\\
{\bf Acknowledgements}\\ 
This work has been supported by the Spanish Research Agency (Agencia Estatal de Investigaci{\' o}n) through the grant IFT Centro de Excelencia Severo Ochoa SEV-2016-0597, and the Spanish Government grants FPA2015-65480-P, FPA2016-78022-P.  The work of GC was supported by Funda{\c c}{\~ a}o para a Ci{\^ e}ncia e a Tecnologia (Portugal) under project  CERN/FIS-PAR/0022/2017 and contract `Investigador FCT - Individual Call/03216/2017'. This project has received funding from the European Union’s Horizon 2020 research and innovation programme under grant agreement No 824093.

\end{document}